\documentclass[sigconf]{acmart}

\AtBeginDocument{%
  \providecommand\BibTeX{{%
    \normalfont B\kern-0.5em{\scshape i\kern-0.25em b}\kern-0.8em\TeX}}}
\usepackage{graphics}
\usepackage{multirow}
\usepackage{graphicx}
\usepackage{subcaption}

\copyrightyear{2025}
\acmYear{2025}
\setcopyright{cc}
\setcctype{by-nc-nd}
\acmConference[CHI '25]{CHI Conference on Human Factors in Computing
Systems}{April 26-May 1, 2025}{Yokohama, Japan}
\acmBooktitle{CHI Conference on Human Factors in Computing Systems (CHI
'25), April 26-May 1, 2025, Yokohama,
Japan}\acmDOI{10.1145/3706598.3713105}
\acmISBN{ 979-8-4007-1394-1/25/04}

\begin{document}

\title[Infrastructures for Inspiration]{Infrastructures for Inspiration: The Routine Construction of Creative Identity and Inspiration}

\author{Ellen Simpson}
\email{ellen.simpson@virginia.edu}
\orcid{0000-0003-0387-7329}
\affiliation{%
 \institution{University of Virginia}
 \streetaddress{1919 Ivy Road, P.O. Box 400249}
 \city{Charlottesville}
 \state{Virginia}
  \country{USA}
 \postcode{22903}
 }
\author{Bryan Semaan}
\email{bryan.semaan@colorado.edu}
\orcid{0000-0003-1151-2389}
\affiliation{%
 \institution{University of Colorado}
 \streetaddress{1045 18th Street, Campus Box 315}
 \city{Boulder}
 \state{Colorado}
  \country{USA}
 \postcode{80309-0315}
 }

\renewcommand{\shortauthors}{Simpson \& Semaan}

\begin{abstract}
Online, visual artists have more places than ever to routinely share their creative work and connect with other artists. These interactions support the routine enactment of creative identity in artists and provide inspirational opportunities for artists. As creative work shifts online, interactions between artists and routines around how these artists get inspired to do creative work are mediated by and through the logics of the online platforms where they take place. In an interview study of 22 artists, this paper explores the interplay between the development of artists' creative identities and the, at times, contradictory practices they have around getting inspired. We find platforms which support the disciplined practice of creative work while supporting spontaneous moments of inspiration, play an increasing role in passive approaches to searching for inspiration, and foster numerous small community spaces for artists to negotiate their creative identities. We discuss how platforms can better support and embed mechanisms for inspiration into their infrastructures into their design and platform policy.
\end{abstract}

\ccsdesc[500]{Human-centered computing~Empirical studies in collaborative and social computing}
\ccsdesc[500]{Human-centered computing~Empirical studies in HCI}

\keywords{inspiration, infrastructure, creative identity, art, artists, online communities}

\maketitle

\section{Introduction}
    With the growth of online community spaces, today's artists and other creatives have more opportunities and creative spaces than ever before to not only do creative work, but also to share it and interact with the creative work of others. These online spaces are not always good for artists, however, as their presence on these online platforms is precarious \cite{duffy2021nested} and subject to platform governance structures \cite{bishop2019managing, bishop2020algorithmic, ma2021advertiser, riccio2024exposed}. As the communicative norms of platforms are embedded into platform design and policies, many artists find themselves being nudged toward a kind of homogeneity, or "influencer creep" during their routine interactions with these platforms \cite{bishop2023influencer}. Despite the challenging landscape of these online creative spaces that are increasingly mediating people's routine creative experiences and identities, the infrastructures of these spaces allow for the routine enactment and realization of people's creative goals and artistic expression \cite{simpson2023rethinking}. There is less discussion, however, about the interplay of the routines around getting inspired, creative practice, and the enactment of creative identity on these platforms. \par

    HCI researchers have long been interested in art and creative work; exploring DIY, maker and craft communities of practice \cite{JonesHandSpinning2024, Vyasaltrusism2019, EmersonShared2024, frich2018twenty, AndersonShredding2022} as both a subject as well as a means of research inquiry \cite{friske2020entangling, RomeroWoven2024}. Another line of inquiry focuses on creative tool development - where the focus is on augmenting creative practice at various stages of the creative process \cite{WanIdeation2023, hwangideabot2021, frich2019mapping, karimi2020creative}. While many of these tools are targeted at assisting in the ideation process (e.g., brainstorming with digital mood boards \cite{WanIdeation2023, lucero2012framing}), considerably fewer focus on the process of inspiration as it ties to the enactment of an artist’s creative identity. Ideation, an interactive process that is sometimes collaborative with human or non-human entities \cite{LinCollaborativeIdeation2020}, does not occupy the same space in the creative process as inspiration. \textbf{Inspiration is an enhancement of cognitive functions, such as divergent thinking or concept blending, that leads to increased idea generation \cite{desai2024psychology}}. One must be inspired to ideate, and ideation without inspiration is challenging. Inspiration breathes life into, as well as animates, the mind to produce new ideas that would not otherwise come about \cite{hymer1990inspiration} - it is the action that comes \textit{before} the creative practice \cite{hoppe2022before}, and, as Weber suggests, inspiration occurs when it pleases, not so much when it pleases us \cite{weber1919science}. \par

    Doing creative work relies on a variety of routine practices; repeated and recognizable patterns of interdependent action carried out by multiple actors \cite{feldman2000organizational, pentland2012dynamics}. While some scholars argue that routines and creativity are diametrically opposed \cite{amabile1999changes}, others suggest that routines allow the possibility of the enactment of novel, creative ideas \cite{feldman2016beyond, sonenshein2016routines}. An integral routine of creative work involves the often contradictory practices of getting inspired. This paper draws on Hymer's \cite{hymer1990inspiration} contradictions of inspiration, as a useful lens through which to interpret the routine processes and practices of inspiration. In these, one must be disciplined in one's creative practice, but also spontaneous enough to react to a chance encounter with something inspirational; where one must be mindful the inspirational potential of one's surroundings, but still able to mindlessly engage with various spaces for chance encounters with inspirational objects; and how one must actively search for inspiration, but also have the control to wait for inspiration to come to them\cite{hymer1990inspiration}. Online, these practices are mediated by and through a platform's infrastructures, as is the interplay between the routines of getting inspired and the enactment of an artist's creative identity. Creative identity is collectively negotiated, a single person cannot determine their creative identity - or what creativity means in any particular context - it must be understood with, by, and through routine interactions with creative \textit{others} \cite{gluaveanu2014creativity}. \par

    This paper explores the contradictory behaviors of inspiration as they are mediated by online platforms through an interview study with 22 visual artists. We find that online platforms play a role in the development and enactment of artist’s creative identities through a series of relationships with necessary others - the assemblage of entities required for inspiration (e.g., creative peers, recommender systems). We discuss the role of online platforms in supporting creative identity development, contributing design recommendations to better support the mediation of inspiration and creative identity development by online platforms. \par

\section{Related Work}

\subsection{The Routine Infrastructures of Creative Identity}

    To enact one's creative identity, or any identity for that matter, people rely on routines. Routines are recognizable patterns of action or behavior that are carried out by one or multiple actors within a specific context \cite{feldman2000organizational}, and are assemblages of sociomaterial configurations people and artifacts (e.g., tools, procedures, technologies) \cite{pentland2012dynamics, shelby2024creative, latour2007reassembling}. People have agency to adapt their routines or to create new ones as need be \cite{pentland2012dynamics}. Having the agency to adapt is key to the foundational routine of building and rebuilding a coherent and rewarding sense of identity \cite{giddens1991modernity}. A strong sense of self-identity, which is how a person thinks about themselves socially or physically \cite{gecas1982self}, can give people a deep sense of security in their everyday lives \cite{ibarra2010identity}. This sense of security emerges when routines are continuous and predictable, a state of ontological security, emerging from the "routine project of the self" \cite{giddens1991modernity}. The flexibility in the routine project of the self comes from people's abilities to consciously or unconsciously use inferences from the past to anticipate a future \cite{giddens1991modernity}. Thus, identity is a routine personal and social undertaking—as one must routinely interact with the world and reflect on the impacts of those interactions in their routine project of the self. 
    
    The routines of our everyday lives are enacted on, through, and within larger societal systems. The foundations of these systems are known as infrastructures, and they support the large scale-systems that society relies upon to routinely function \cite{edwards2003infrastructure}. Infrastructure can be anything from large-scale highway systems to information and communication technologies (e.g., social media platforms) in how they support routine societal function \cite{hanseth2010design}. Infrastructure is defined in use – as they are entwined with human social practice as relational systems that take on meaning or changes in meaning in a continually negotiated way depending on the social practice taking place and the actors involved \cite{star1996steps}. Importantly, while humans are a part of the construction of the social meaning of infrastructures, they can be infrastructures, functioning as a combination of both known and unknown entities that animate a particular system \cite{LeeDourishMark2006}. In this sense infrastructures are sociotechnical, meaning that they both shape and are shaped by social practices built around and with them \cite{edwards2003infrastructure, star1996steps}. As with any large system, infrastructures have many interconnected parts that weave themselves seamlessly into the fabric of society and go unnoticed most of the time \cite{star1996steps}. \par

    While only one small aspect of ourselves, being \textit{a creative person} is also something that is produced through routines. Creative identity is a "representational project engaging the self in dialogue with multiple others about the meaning of creativity as constructed in societal discourses" \cite{gluaveanu2014creativity}. Creative identity cannot be understood by the actions of individuals alone, but rather relationships and connections between the self and others as they develop a shared notion of creativity \cite{gluaveanu2014creativity}. To have a creative identity, a person must do creative work, present that creative work to others, and have others also deem that work to be creative. One must be flexible in the routine project of the self \cite{giddens1991modernity}, and creative identities are no different. Drawing ontological security around creative identity requires flexibility in how the artist constructs knowledge about the world and about themselves as creative people. \par
   
    Creative identities are supported by human infrastructures. There are people that do the work required to animate the physical and digital infrastructures where creative work is shared, as well as the electric and network infrastructures that mediate the routine presentation of one's self as a creative person. Often, artists do not know who these human infrastructures are, but their work is vital to the continued ability of creatives to routinely express themselves as creative people. \par

    Yet, sometimes infrastructures do not work in the way that they are intended to, and sometimes they break down \cite{star1996steps}, which can become chronic in certain circumstances \cite{Semaan2019}. For artists, infrastructural breakdowns may come from how their needs and values may not match the ways infrastructures are designed \cite{simpson2023captions, simpson2023rethinking}, as infrastructures are not value-neutral. The human actors that build, maintain, and repair infrastructures embed their values, norms, and biases into them through this routine work \cite{bowker1994information, bowker2000sorting, JacksonValuesInrepair}. As they are embedded into infrastructures, these values can be at the heart of the routine sources of disruptions in artists' everyday lives and routines \cite{Semaan2019}. In some cases, value misalignment between infrastructure designers and users can result in incomplete infrastructure---an infrastructure that does not meet the needs of those who depend upon it to enact their routines, which is a common concern for artists \cite{simpson2023captions}. Infrastructural breakdowns may surface the underlying ways these infrastructures support the routine development and expression of creative identity \cite{gluaveanu2014creativity} - such as how people routinely draw on online spaces like Instagram to be inspired by the creative work of others or how designers use mood boards to frame or direct their design process \cite{lucero2012framing}. When online spaces fail to meet artists’ needs, they can leave people at a loss of where to find the people and creative objects necessary to feel inspired \cite{lucero2012framing} and to negotiate their creative identity with others \cite{hart1998inspiration, gluaveanu2014creativity}.  \par

\subsection{The Sociomaterial Foundations of Inspiration}
    The word \textit{inspiration} has its root in the Latin \textit{inspirare} - which means to breathe on or into, or to animate the soul \cite{hymer1990inspiration}. During the course of our routine encounters with the world, we will, on occasion, develop "intense object relationships" with \textbf{necessary others} that are the seeds of creative products or ideas that would not otherwise come about \cite{hymer1990inspiration}. Put another way, when we, the subject of these interactions, are inspired, we are putting ourselves in direct relationship with a necessary other, which could be anything. These relationships are contradictory in nature, as inspiration requires both "discipline and spontaneity, mindlessness and mindfulness, receptive waiting and active searching" to come into being \cite{hymer1990inspiration}. Similar to how routines are patterns of action that everyone enacts in a slightly different way than everybody else; inspirational objects and the transformational relations they produce, are not static, but rather emerge in slightly different ways each time the sociomaterial relationships and particular conditions that evoke these creative products or concepts to come into alignment \cite{feldman2000organizational, pentland2012dynamics, rudnicki2021ideas, hymer1990inspiration}.\par

    For the purposes of this paper, we understand sociomateriality as the entanglement of people and objects during the routine actions of individuals or collectives within organizations \cite{orlikowski2007sociomaterial, cook1999bridging}, such as in how open-concept offices often have an organization's management team sitting in corners or around the edges of a collective workspace to enforce sociomaterial control over interactions between their team and others \cite{perriton2023constitutive}, which can only emerge through an entanglement of physical space, humans, and technology objects. The intertwining of humans and objects is situational, meaning that it is produced and reproduced differently depending on the context within which the practices are taking place. Inspiration emerges from the sociomaterial relationships between people and objects that we encounter as we go about our everyday lives. But, importantly, while an object may shape the practice of an individual in a specific organizational context, it is just one thing out of many things that exists within that context. This object shapes the practice of that individual in that particular context, which, in turn, shapes the object itself \cite{oraghallaigh2017sociomateriality}. In a sociomaterial context, a necessary object may be inspirational to one person, but may not be inspirational to another -- yet when that one person acts on that inspiration, they in turn shape the necessary object, which may lead to it becoming inspirational to someone else. \par

    Sharon Hymer \cite{hymer1990inspiration} points out four key relationships that are particularly generative of inspiration: relationships with the divine, with inanimate objects (e.g., music or nature), the secular (e.g., mentors, teachers), and the self. These relationships emerge differently for different people, and what is inspirational to one person may not be inspirational to another - or it may not be inspirational in the same way. For artists, these encounters with inspirational objects are happening as the result of creative work taking place increasingly online. Prior work has shown that encountering the creative work of others has led to what is known as "divergent thinking" -- the free-forming of new ideas that branch off from the original idea or concept \cite{GallagherIdeation2017} and that attempting to copy or recreate the creative work of others allows for transformation of the original idea into something new \cite{okada2017imitation}. Graphic designers, for example, seek out visual information--such as the creative work of others--to inform their personal development as well as to capture certain aesthetics or ideas as a part of their inspiration and ideation process \cite{laing2015study}; or will put together moodboards (usually of other people's art or photographs) that frame, direct or otherwise inspire their design process \cite{lucero2012framing}. This prior work demonstrates that artists are inspired by their encounters with, deliberate search for, and transformation into inspirational tools like moodboards; the creative work of others.\par

    At the heart of many of these encounters is technology, which facilitates the creative process and management of creative ideas on both an individual and collaborative level \cite{Rosselli2024Ideas}. Online, these spaces and encounters are more plentiful than ever before and are increasingly facilitated and impeded by platform infrastructures \cite{simpson2023rethinking,simpson2023captions,Rosselli2024Ideas}, meaning that the artist who creates and shares their work online is constantly bombarded by the potential for inspiration that stems from the creative work of others. One may be deliberate in seeking out visual information to help with the creative process \cite{laing2015study,hymer1990inspiration}, but one could also stumble on an inspirational object without actually meaning to look for one. How these spaces are designed and the infrastructural elements that facilitate these contradictory inspirational encounters can place the creative self - a person's \textit{creative identity} - in flux. The constant search for inspiration becomes a matter of routine engagement with one's creative identity as it relates to the creative expression and identities of others, while it also is, increasingly reliant and entangled with technical tools and their infrastructures. 

    While creative practice is a matter of routine for many artists, it is also a routine that is pulled in contradictory directions when it comes to how inspiration emerges. Inspiration emerges along three key contradictions: one must be disciplined, as well as able to be spontaneous to become inspired; one must always be mindless, but also mindful of when inspiration may strike; and one must be actively searching for inspiration, while also receptively waiting for a chance encounter with an inspirational object \cite{hymer1990inspiration}. We adopt Hymer's \cite{hymer1990inspiration} contradictions as a lens by which to understand the role inspiration plays in the enactment and realization of creative identities. Below, we detail the contradictions, ground them in the existing literature, and provide definitions. 

\begin{itemize}

\item \textbf{Discipline and Spontaneity} are centered around creative practice itself and the inspiration that emerges from the practice of doing art. According to Hymer, "what appears to be a serendipitous thought or discovery [...] derives from both flashes of quick impressions and from the slower, more painstaking analytic work that precedes and follows such flashes" \cite[~p.30]{hymer1990inspiration}. For our purposes, discipline surrounds the routine practice of creative work that comes both before and after the spontaneous, serendipitous, moment of inspiration. Self-discipline is the routine enactment of creative practice through the discipline to do creative work even when struggling with creative blocks \cite{gallay2013understanding}, by drawing on peer support \cite{wallace1987using} and engaging in critique and feedback sessions \cite{gluaveanu2014creativity} that negotiate meaning of creativity and artist identity. Spontaneity, conversely, is the ability to capitalize on the convergence of inspirational objects and necessary others that produce new combinations of possibilities that inspire \cite{derond2014structure} and then being open and able to do that creative work \cite{thrash2014psychology}. 

\item \textbf{Mindfulness and Mindlessness}, according to Hymer are are framed around artistic awareness and engagement with the world \cite{hymer1990inspiration}. These contradictions are embodied in the creative objects produced and the collective negotiation of creativity and creative identity that online platforms facilitate \cite{gluaveanu2014creativity, simpson2023rethinking}, as well as the tools that we use to create potential inspiration \cite{Rosselli2024Ideas,laing2015study,lucero2012framing}. Mindfulness speaks to the awareness of the inspirational elements of one’s environment and the openness to documenting inspirational objects when they are encountered for further reference. Mindlessness, in contrast, speaks to the routine, yet mindless, use of online platforms to engage with the creative work of others and to become inspired to do creative work. We note that a key element of these routine encounters is engagement with various recommender systems, which mediate the artists’ encounters with the creative work of others \cite{simpson2022tame}. 

\item \textbf{Active Searching and Receptive Waiting} are focused on a person's intentionality in searching for inspiration. Hymer describes this contradiction as being both aware of how one has routinely gone about inspirational practices in the past (e.g., through deliberate routine), but also the "spontaneity and receptivity to surprise elements which enter into a transformational experience" \cite[~p.31]{hymer1990inspiration}. While actively searching for inspiration is a routine practice for many creative people \cite{hill2016searching}, receptive waiting is about browsing - or, “a search, hopefully serendipitous [...] which might contribute the fact or idea needed in some intellectual effort” \cite[~p.4]{morse1970browsing}. Here, an element of control that must be exercised to create the possibilities of chance encounters that could be considered “serendipitous” - thus, receptive waiting  \cite{rice2001accessing, foster2003serendipity}.

\end{itemize}

    Using this framework, we describe how the online platforms where many artists are encountering inspirational objects are supporting - and not supporting - the articulation of their artistic identities. In the next section, we discuss our method. \par

\section{Method}

\subsection{Participants and Recruitment}
    We recruited participants from a wide variety of channels, drawing on in person social networks, social media, snowball sampling, and convenience sampling to allow for a wide range of participants engaged in creative work. A recruitment flier created and shared in coffee shops and artists spaces across the states of Colorado, Oregon, Washington, and Southeast Alaska, on the personal Twitter, Tumblr, and Facebook accounts of the first and second authors, and in several artist Discord communities the first author moderates. The recruitment flier included a link to an interest survey hosted on our university's Qualtrics. The initial question on the survey included front matter about consenting to be a part of the research project, as well as contact information for our university's Institutional Review Board. The survey gathered basic demographic information about the participants (e.g., age, gender expression), and contact information. 14 responses were received on Qualtrics, and a further five were received on Discord. \par

    Next, the research team contacted five artists within their social network who live in diverse settings (urban, suburban, and rural), following a convenience sampling approach similar to Ma and colleagues \cite{ma2023multi}'s recruitment strategy for content creators. Using the snowball recruitment method \cite{biernacki1981snowball}, following our interviews we asked them to share our recruitment flier with their networks. The research team was contacted by two additional working artists using this method. \par  

    Finally, after conducting the first ten interviews of this study, the first author explored other opportunities of direct solicitation for participants. We joined several artist Reddit communities and identified people asking questions about sharing their art online, and people replying to these posts. The first author used their personal account to individually message 13 users participating in these discussions expressing interest and referencing their post directly. If the message was responded to, we would reply with a link to the study recruitment survey. We drew on first author’s personal Instagram account to message the first 18 video makers who had used a trending sound related to loving art and independent artists and whose videos had under 1000 views. We selected the under 1000 views criteria after comparing those views to other views using the trending sound. The research team, following previous HCI research \cite{foryouforyou, simpson2023captions}, determined that these video makers may be more likely to respond to potential research inquiries. As with Reddit, we sent details of the study and and, if the message was responded to, a link to the recruitment survey. \par

\begin{table}[h]
\centering
  \centering
	\begin{tabular}{|l|l|}
    	\hline \textbf{Recruitment Site} & \textbf{Participant(s)}  \\ \hline
    	Personal Social Networks & P1, P7, P8, P13, P15  \\ \hline
    	Extended Social Networks & P11, P16  \\ \hline
    	Discord & \begin{tabular}[c]{@{}l@{}}P2, P3, P4, P5, P09\\ P10, P12\end{tabular}  \\ \hline
    	Twitter & P6, P14 \\ \hline
    	Reddit & P17, P18, P22 \\ \hline    
    	Instagram & P19, P20, P21 \\ \hline   
	\end{tabular}
	\caption{List of Participant Recruitment Sites}
	\label{tab:recruitmentsite}
	\Description{A list of recruitment sites of our  participants. Data Table Follows.}
\vspace{-8pt}
\end{table}

    By triangulating these three recruitment methods, we reached out to 55 potential participants, and ended up successfully recruiting 22 artists, whose recruitment sites are detailed in Table \ref{tab:recruitmentsite}. Participants in this study were diverse in terms of age (18 - 74), locale (41\% Urban, 36\% Suburban, 23\% Rural), gender expression (32\% Men, 36\% Women, 27\% Gender Non-Conforming or Non-Binary, 1 participant [5\%] did not self-identify) and the way they described their art. The participant pool was less diverse in terms of racial diversity, with only 36\% of the participants identifying as Black, Latine/Hispanic, Ashkenazi, or Asian American. While these results are presented in aggregate form, the full participant demographics are included in Appendix A. \par

\subsection{Interviews}
    To best understand the interplay between inspiration and the routine project of the creative self, we conducted 22 semi-structured interviews (mean length: 70 minutes, range: 60-120 minutes) that took a life history approach \cite{wengraf2001qualitative} as we wanted to understand how participants grew as artists over time. Two interviews (P17, P18) were conducted via text on Discord, three (P16, P19, P20) were conducted over the phone, One (P11) was conducted in person, and the remaining (P1 - P10, P12 - P15, P21, P22) interviews were conducted via Zoom. While we conducted interviews through several different means, we find that there was little difference in the resulting findings, echoing \cite{dimond2012qualitative}. With participant consent, we recorded and transcribed the interviews. We used a small grant from our university's graduate student funding body to transcribe poor audio quality interviews (P11, P14, P19, P20) using the transcription service, Rev. \par

    The interview was divided into several parts: First, learning more about the participant’s identity, life history, and how they came to do the art they currently do \cite{wengraf2001qualitative}. Next, participants were asked about their creative work and how and why they share it in various contexts. We paid particular attention to the routine elements of the creative work participants do around sharing their art in particular spaces or on particular platforms, what they share, and whether that art has changed over time. The interview also included questions about places participants go to for social and creative support. The second section of the interview assessed if participants monetized their art in different ways across platforms, as well as their feelings about the monetization of art generally. Capturing these thoughts is important as prior work \cite{bishop2019managing, duffy2021nested} has shown that the visibility of an individual’s user-generated content can impact what they create or share on a particular platform. \par

    Finally, the interview included a optional concept mapping exercise where participants were asked to draw their creative process from inspiration to finished product. In qualitative and social science research, concept maps are often used as a visual expression of meaning that a participant generates which can facilitate data collection \cite{wheeldon2009framing}. In HCI, concept, mind, and cognitive mapping have been used in numerous contexts as an analytical tool (e.g., \cite{devito2018too}) and are good for tracing routine actions. Although only a handful of participants participated fully in the exercise as designed (P2, P5, P6, P9, P10, P12), the conversations across the board during this “please describe your creative process” phase of the interviews were fruitful, and typified how participants thought about their creative processes. \par

\subsection{Analysis}

    Once the interviews were fully transcribed and checked for transcription errors, the first author conducted a round of open coding using an inductive approach based in grounded theory \cite{strauss1990basics}. This approach is commonly used in HCI research \cite{karizat2021algorithmic, britton2019mothers, foryouforyou}. During the initial round of opening coding, the first author assigned tags freely. After tagging five interviews, the first and second authors met weekly to discuss the emergent codes and potential means by which to collapse them. Tags were subsequently collapsed and sorted using MAXQDA, a qualitative analysis software. \par

    Emergent from this narrative was an examination of the creative processes of the participants. Conversations focused on routine creative labor, and how these artists build systems for themselves that drew on available infrastructures for creative support -- particularly around the relational contradictions of getting inspired to do creative work. These findings led us to review literature on inspiration and to read Hymer's analysis on the contradictions of inspiration, which supported our observations and provided a helpful lens through which to reinterpret the results \cite{hymer1990inspiration}. The first author did a secondary pass in a more deductive fashion on the findings to 1) extract where these contradictions were already coded within the data, and 2) to identify further instances of where these contradictions were clearly illustrated, which subsequently went on to inform the narrative flow of this paper. The first and second author met twice during this time to discuss emergent findings and code groupings. What emerged from this was a conversation about how the routine enactment of creative identity was tied up in the contradictions of inspiration, as filtered through the platform infrastructures upon which that enactment took place. \par

\subsection{Reflections and Limitations}
    The first author is a watercolor artist, which helped to build rapport with many of the participants in this study. While we spoke to 22 artists, we did not speak to enough of any particular type of artist to claim comprehensive knowledge of one particular cohort of creative people. To this end, we have taken efforts to not speak to specific artistic forms, but rather creative processes broadly. While there are myriad social, cultural, and historical power dynamics at play when reflecting on what is and is not considered "art" that are contentious along racial, gender, and class lines, our cohort skews white. While race did had little bearing on the discussions at hand, we recognize as a white and Iraqi writing team that our identities may have played some inadvertent role in our interview processes, and collaboration on data analysis to mitigate any potential biases in our analyses. 

\section{Characterizing the Contradictions: The Infrastructuring of Inspiration}
    In this section we explore Hymer's contradictions of inspiration \cite{hymer1990inspiration}. Even though they were not directly asked about getting inspired during our interviews, participants tended to start describing their creative process with how they got inspired to do art. This process was contradictory in nature, as participants often described very intentional processes around \textit{looking} for inspirational objects, while also pointing out that sometimes inspiration just kind of happened to them in a serendipitous moment of interaction with an inspirational \textit{something}. In the first section, we discuss the contradiction of the discipline of repeated creative practice as it interfaces with the spontaneous flash of inspiration \cite{hymer1990inspiration}. We then discuss mindfulness and mindlessness during routine creative practice. Finally, we discuss how one must be always searching for inspiration, but also receptive to the moments where inspiration strikes.  \par

\subsection{Discipline \& Spontaneity: Drawing on Platform Infrastructures for the Inspiration to Do Art}
    In this section, we discuss the contradiction of discipline and spontaneity, focusing on how participants drew on platform infrastructures to do art, find inspiration, and express their creative identity.

\subsubsection{Discipline}
   
    The self-disciplined practice of deciding to do creative work participants described often drew on the infrastructures of online spaces - both human and technical - to help facilitate that practice. Creative identity stems from the social interaction of a creative individual and their peers, allowing for the collective negotiation of their identities as artists as mediated by their creative work \cite{gluaveanu2014creativity}. P3, a freelance artist and illustrator, talked about a feature of a Discord community they belong to:

    \begin{quote}
    “So joining [discord servers], right, there is the Power Hour, right? Which was huge. That was really helpful too, as someone with ADHD, to have an accountability space, right? [...] because we're all talking about creating something in that time. And then in this [discord], they have a channel that is just like art critique, so people will post things that like, I'm trying to do this, does this detail work, how do I do it? I really enjoyed that.”\end{quote}

    Power Hours, according to P3, are a declared period of focused work time that are similar to the Pomodoro technique \cite{cirillo2018pomodoro}, where people state their intentions for work at the start of the hour, and work on that set task until the end of the hour when they return to the channel and share out their progress.  P3 used this designated time deliberately, drawing on Discord’s infrastructures to facilitate focused work time and become inspired to do creative work through social and creative interactions by creating a disciplined accountability space with others. Motivation and openness to do creative work is a form of discipline, which is key to the enactment of a creative identity \cite{thrash2014psychology}, and many participants described relying on personal social networks (P7, P8, P16, P20), professional networks (P2, P19, P20) and people in artistic communities on online platforms like Reddit (P8, P17, P18, P22) or DeviantArt (P10, P15, P17, P18) to find and support that disciplined creative practice. The peer support networks, such as P3's Discord-supported Power Hour, functioned as the necessary human others to both facilitate inspiration and create the discipline required for artists like P3 to articulate and realize their creative identities.

\begin{figure}
	\centering
	\includegraphics[scale=.60]{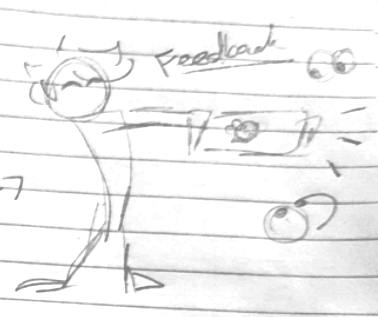}
	\setlength{\belowcaptionskip}{-10pt}
	\caption{P2's illustration of the role getting critique and feedback plays in their creative process. }
	\Description[A hand drawn sketch drawn by P2.]{A hand drawn image by Participant 2 showing a figure holding up a piece of paper with eyeball emojis surrounding the figure with the text reading 'feedback.'}
	\label{fig:P2_Feedback}
\end{figure}

    For other participants, inspiration emerged through their routine use of social chat and streaming platforms like Discord (P1-P6, P9, P10, P12, P13, P14, P17) or Twitch (P1, P9) or YouTube (P9). In these spaces, participants spoke of how they collaborated with peers and relied on that interaction to inspire their routine creative work. P2, a 21 year old animator, told us about how they use Twitch to get feedback, illustrated in Figure \ref{fig:P2_Feedback}:

    \begin{quote}
    "There's a couple of Twitch streams, like a few people that taught at the school that I went to, and they'll be like, if you're on and are open to feedback, send your work and we'll leave you notes kind of thing. [...] Like it's really like 30 people in a stream at any given time. But those are super helpful [...] So there are people out there who will just want to take a look at your work and leave with thoughts on it."
    \end{quote}

    P2's experiences with these Twitch channels, and their consistent reliance on them as a key part of their routine creative process, has provided them with a way to feel inspired not just from the critique they receive, but also in their interaction with creative people who value and see their art and want to help them improve as an artist. In most cases, participants felt that the online platforms and creative spaces they went to were very supportive of their continued project of their creative selves and continued discipline to do creative work. \par

    Participants also extolled the virtues of the online platforms used and the social connections maintained by and through these spaces as helping them to be disciplined in their engagement with and enactment of their creative identities, and in inspiring them to do creative work. Discord was not unique in this respect, with other participants (P1, P2, P6, and P9) discussed doing live art streams on video streaming platforms like Twitch (P1, P2, P6, and P9), YouTube (P9), and sharing videos of themselves doing art on Instagram (P18, P19, P20, P22) and TikTok (P22) as a part of being disciplined in their creative process. These were done alone in some cases, but many participants also framed this as a disciplined way to engage in creative work even when they were not inspired. \par

    The interplay between creative people and the social connections that platform infrastructures supported, were inspirational to many of participants as they helped them to foster and feel ontologically secure in their creative identities. These social connections allowed participants to achieve their creative goals, either through discipline and accountability to get inspired to do their creative work (P1, P2, P3, P5, P6, P9, P10, P14, P15, P20, P21), or to work with an inspirational cohort of peers that both inspired them and pushed them creatively (P1 - P5, P8 - P10, P12 - P15, P20, P21). The social aspect of inspiration, where ideas generate from people’s everyday interactions with friends, peers, and audiences, known or unknown, is a key element of the everyday routines participants describe around their creative processes. These spaces further foster the development of a secure sense of creative identity, as these artists are interacting with peers that are functioning as the necessary others within creative spaces.\par

\subsubsection{Spontaneity}
    Participants discussed how inspiration emerged through their everyday interaction with others - particularly friends or acquaintances within platforms that have a strong creative component to the people who gather there. While the artists we spoke to were very self-disciplined in their approaches to seeking creative feedback and in their doing creative work around and with others, sometimes the interactions with these creative people allowed for the spontaneous emergence of inspiration. Spontaneity, here, is the organic inspiration and interaction that emerges from the convergence of creative people who play off of each other, essentially functioning as human infrastructures for creativity that support and allow for the routine emergence of inspiration and organic support of creative identity. Thirteen of twenty-two participants cited Discord as a place that provided participants space to discuss common interests and creative ideas—collectively negotiating the meaning of creativity with fellow creatives peers—which often would lead to moments of inspiration that were subsequently supported by the communities contained within these online spaces. \par

    For example, P10 described an experience of getting really inspired at the start of a new Dungeons and Dragons (D\&D) campaign that was played with a disbursed group of internet friends via a designated Discord server. According to P10, the inspiration to do this art emerged on a whim as players described their characters during the first session of the game. She explains:

    \begin{quote}
    "It was like that first [...] week that we started playing D\&D, I had drawn like four different characters, full bodied, or like half bodied, with outfits and shit. And I was just like, really motivated to do it." \end{quote}

    P10’s new D\&D campaign inspired her to create a series of original art pieces, and to produce them faster than usual. Discord served as a social platform for P10 not only to share art and to be validated in her creative identity, but also for a space where she found inspiration and encouragement in her art. Discord’s infrastructures supported the routine social interactions she had with her peers around this new D\&D campaign, allowing her to be both inspired by the new creative setting she and her friends were building, but also inspired to create because of it. What is more, P10 was not being disciplined when she went into these spaces – she did not have the intent of purposefully searching for inspiration while she used Discord, yet the convergence of creative people on a platform whose infrastructures supposed the ready exchange of ideas and conversation through its chat-based design allowed for inspiration to strike for P10 in a spontaneous way. \par

    \subsubsection{Mindful Looking}
\begin{figure}
\centering
\begin{minipage}{.45\textwidth}
  \centering
  \includegraphics[width=.8\linewidth]{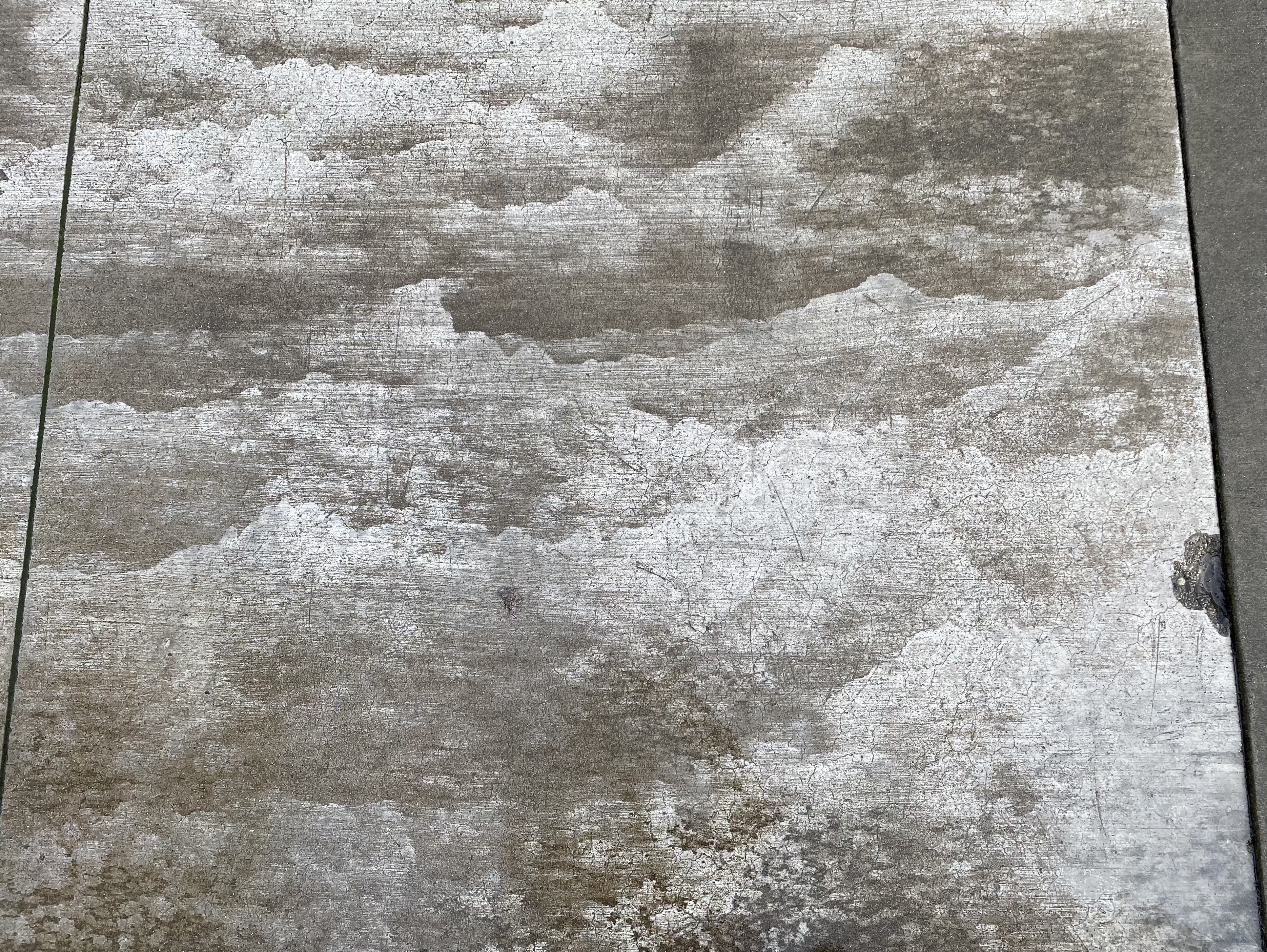}
  \captionof{figure}{A Photograph of a Sidewalk P7 Took.}
         \Description[A photograph of a sidewalk.]{A photograph of an expanse of sidewalk that P7 took.}
  \label{fig:P7}
\end{minipage}
\end{figure}

    This experience was common across all participants in this study. Creating art – and creative drive and inspiration – was commonly framed as a social interaction with another human, or a platform that hosted other people’s creative work, or as watching a particularly interesting cartoon or movie with a Discord server full of friends. Creating together with others is an important unstructured element of creativity that is supported by the routines participants have around becoming inspired which draw on online spaces to emerge. P4, a 29 year old illustrator, told us that sometimes Discord spaces allowed them to break through creative blocks and a lack of motivation to do art. They explained:

    \begin{quote}
    “I have a lot of artists block most of the time and sometimes[...] I will ask a friend on the server or just like a friend who also does art [to] like, give me a prompt.” 
    \end{quote}

    Repurposing Discord into places that inspire their creative drive and connect themselves to necessary creative peers and mentors both allows for inspiration to spontaneously emerge from these interactions while also validating the individual artist's creative identity. We also saw this on Instagram, where the connections were driven not by intentional connection to creative peers and mentors, but rather by algorithmic content recommendation of a person's art to random strangers. For example, P19 described feeling both very validated and inspired by the feedback she got from her Instagram audience as she worked her way through a 100-day bookmaking challenge. All told, these platform infrastructures were useful in supporting both creative identities and allowing for spontaneous inspiration to emerge, particularly if one chose to engage with curated communities and spaces that were directly geared toward creative support and social connection.  \par

\subsection{Mindful Looking \& Mindless Browsing: Being Open to the Possibilities of Inspiration}

    In this section, we discuss the contradiction of mindfulness and mindlessness. According to Hymer, this contradiction is framed around awareness and engagement with the broader world \cite{hymer1990inspiration}. Therefore, we focus on how participants described being open to inspiration, how they encountered inspirational objects, and how that openness informed their expression of their creative identities.

\begin{figure}
\centering
\begin{minipage}{.45\textwidth}
  \centering
  \includegraphics[width=.8\linewidth]{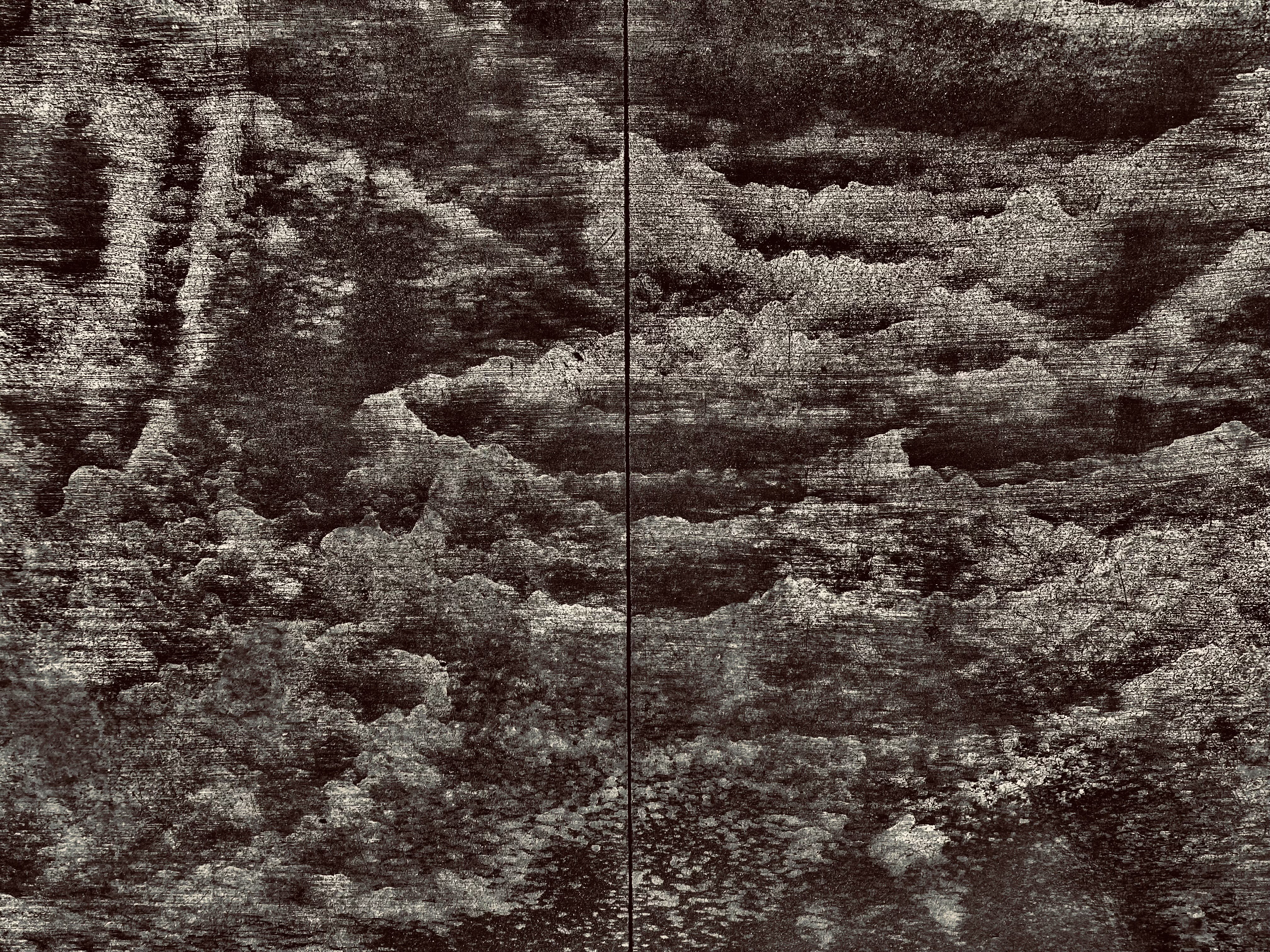}
  \captionof{figure}{A recolored version of the photograph P7 Took.}
    \Description[A colorized photograph of a sidewalk.]{A recolored photograph of an expanse of sidewalk that P7 took.}
  \label{fig:P7_Color}
\end{minipage}
\end{figure}
    The artists in our cohort spoke at length about how being able to document inspirational objects from nature or their environment when they were encountered required a degree of mindfulness -- \textit{mindful looking}. Participants described using digital tools to gather such inspirational objects, such as image collections (i.e., Pinterest boards or mood boards) (P9, P10, P13, P15, P19, P22), while others described using physical tools. P7, a 65 year old sculptor (in wood), described a constant, mindful, looking for inspiration as he went about his everyday routines. He explains,

    \begin{quote}
    “I make use of whatever my eyes and brain find appealing. A good case in point is I was walking [...] down the street, there was this section of pavement -- sidewalk, cement, where the metal posts and everything that were up the street from it, had sloughed off chemicals and colored the cement itself. And I took a bunch of photographs of that. I find a lot of inspiration, [...] things just appeal to my eye and my sense of art, mostly from the natural world. [...] That section of sidewalk was absolutely stunning. To me anyway.”
    \end{quote}

    Figure \ref{fig:P7} is the raw version of one of the photographs that P7 took of the section of sidewalk he references. The seeming randomness of this encounter with an object that was inspiring to P7 in highlights how mindfulness plays a key role in the inspirational process, as he was able to snap this photograph and document something inspirational as soon as he noticed it. P7's ability to snap a photograph of the sidewalk, and later manipulate it into an image he shared with friends (see Figure \ref{fig:P7_Color}), required him to draw on both the physical (e.g., his cellphone), as well as the wireless cellular network infrastructure to share this photograph and it's subsequent manipulation with friends and peers as means by which of having his creative identity valued – as he was able to express the creative and artistic merit of what he saw to others and have them, in turn, validate his assessment of its creativity. The routine nature of P7's mindful engagement with potentially inspirational elements points to how, without the technology he uses to document such encounters, he would not be able to easily capture, and return to reference such inspirational objects and form the intense object relationships that allow for the emergence of new ideas \cite{hymer1990inspiration}. A key characteristic for these sorts of experience of routine, yet mindful, engagements with the world is the openness to the potential of inspiration that may come through these engagements by routinely drawing on the world around them and documenting the inspirational objects they encounter through reliance of an ecosystem of sociotechnical tools. \par

\subsubsection{Mindless Scrolling}

    Despite mindful engagement with the world, there is a certain mindless nature that characterizes encounters with inspirational objects--an openness to the potential of inspiration, or an artistic awareness of the world. Participants described how they would routinely browse specific apps, such as Instagram or Pinterest, drawing algorithmically-curated feeds of content to get inspired by the creative work of others. When browsing, recent research has noted that element of control that must be exercised to create the possibilities of chance encounters that could be considered “serendipitous” \cite{rice2001accessing, foster2003serendipity}. The integration of routine use into otherwise mindless browsing behaviors on apps like Pinterest or Instagram with a specific intent, but no key goal in mind is an example of this control \cite{foster2003serendipity}. Each moment of inspiration participants described was very individual, differing in the nature of the encounter and what about that moment was inspiring.\par

    For example, P13, a 33 year old photographer and crafter, explained that she gets inspiration from other people’s art through her casual browsing of the image-sharing app, Pinterest:
    \begin{quote}
    “Pinterest is a big one for me. I just kind of get other ideas from other people, but not necessarily using what they do, but just see what other kinds of ideas are out there.” \end{quote}

    P13’s use of Pinterest to observe the creative work of others was supported by infrastructural elements of Pinterest, specifically its content recommendation algorithm and versatility as a place to store multiple different collections of creative ideas, where it functions similar to moodboards\cite{lucero2012framing}. This was part of P13’s everyday use of Pinterest, where she described herself browsing in an unintentional, mindless way, during her free time with the openness to finding new ideas to try. \par

    While browsing an app full of creative ideas does communicate at least some intention, even if the practice was mindless, any participants (P1 - P6, P9, P10, P12, P14, P15, P18) discussed consuming media that they enjoyed, or turning to books (P3, P9) or music specifically (P2) to get inspired. P9 explains:

    \begin{quote}
    "And some other times I'm not 100\% sure what I even want. That's where it will result in a lot of mindless searching, browsing, flipping through books, putting on movies, shows, you name it, and just kind of waiting for something to just light up the blue light bulb in my head that 'Oh, yes, this is what I could do.'" \end{quote}

    P9 draws on the platforms that they have integrated into their creative routines in addition to books, films, and other media. For P9, the inspiration they find through their routine, unstructured browsing of the online platforms was similar to the inspiration found through books, films, and other media. Yet P9 also is exercising control over this unstructured browsing. They are not looking for anything in particular, but they are looking for \textit{something} that they find inspiring. The routines they have around looking for that \textit{something} are supported by the infrastructural elements of the platforms they go to look for them on – for example, browsing other people’s art on Pinterest or scrolling through Instagram.\par

   Yet sometimes these interactions cannot be exclusively mindless, because there is the potential to see things that an artist may not wish to see, or things that may be demotivationing, rather than inspirational. P10, a 28 year old hobby artist, struggles to just mindlessly look at Instagram, because she could see content she does not want to see:

    \begin{quote}
    “I don't want to be mean when I say this, but like if I'm on Instagram and I see cute art and I click it and then I see you know, a straight Christian woman trying to draw something or other. [...] It's always a weird, jarring feeling because I feel like, ya know, most of [the artists P10 engages with on Instagram] are some shade or queer.”
    \end{quote}

    P10 tends to look at art that is reminiscent of her art style, and is largely queer as she herself is queer. Yet, despite this clear interaction with content that should have a discernible series of measurable types \cite{cheney2018wearedata}, Instagram repeatedly shows P10 art by people who may fundamentally disagree with not only her creative interests, but also do not believe that people who are homosexual or queer are entitled to fundamental human rights. P10 went on to say that she did not enjoy Instagram, instead favoring Tumblr, Twitter, and Discord as she has more control in those spaces. Not being able to be truly mindless means that P10 cannot be fully open to the possibilities of inspiration, rather she must exercise more control in an online setting. This tension demonstrates how online platforms cannot, for many people whose creative identities are intertwined with other identities they may have, truly support mindless engagement with the platform in a way that allows the organic emergence of encounters with inspirational objects. \par

\subsection{Active Searching \& Receptive Waiting: Searching for Specifics vs. Encountering Parts of a Larger Whole}
    Finally, we explore the role of intentionality when reflecting on how participants engaged in searching for inspiration as a matter of routine. 

\subsubsection{Active Searching}

    While many participants described routinely mindlessly engaging with online platforms as a part of their creative process, others described routinely searching for inspirational objects. Active searching draws on available infrastructural elements like embedded search and relies on the discoverability of inspirational objects that others produce. Participants (P5, P8, P9, P10, P12, P13, P14, P16, P19, and P22) pointed to their routine use of search engines and the embedded search engines within online platforms to find inspiration for their art. These search queries ranged from what to draw, to how to do specific kinds of art, to searching for elements of the creative work of others to become a part of a broader creative vision. \par

    For example, P22, a 37 year old multimedia artist, explained how she approached the question of what kinds of subjects and artistic styles she should produce and sell in her Etsy shop:

    \begin{quote}"I Google, 'what kind of art do people buy?' And it's like landscapes, abstracts, nudes, and there's a list. And then being like, okay, so I need to learn how to make some of the things in these top categories."\end{quote}

    For P22, searching for inspiration really comes down to a pragmatic decision about what kinds of art she can produce and have commercial success in producing. The deliberate nature of how she goes about getting inspired to do certain kinds of art -- this requires discipline, as well as critical engagement of what, amongst the results, will P22 be able to feasibly produce. \par

    Sometimes search results do not align with platform norms and values, however, which can lead to potential conflicts between the artist and the platform. P22 told us about how she found that some of her art did not meet community standards, and therefore was removed from search/discoverability:

    \begin{quote}"And TikTok too [...] even though it's art and it's supposed to be allowed, they'll remove stuff sometimes and they'll flag your account if you post nude art on there."\end{quote}

    This complaint was echoed by P17, a 29 year old digital artist, who discussed his experiences with Instagram’s moderation algorithms as also negatively impacting what he could share with others - and by extension the art he could freely be inspired to do\footnote{P17 was interviewed via text, which has been reproduced below with only correction for typographical errors}:

    \begin{quote}“[T]he algorithm has changed 2 times since I joined negatively affecting my page in more ways than my usual social-awkwardness does, in one they nuked reach of images in favor of video and the next the \# basically became meaningless if you had a "but this account does X" in your profile, my "but X" is sometimes literal depictions of buts and breasts xD”\end{quote}

    P17 is one of several participants who draws adult content (P6, P9, P17, P18, P22). While his brushes with Instagram’s content moderation algorithms make sense from the perspective of Instagram’s policy on nudity and not-safe-for-work images and videos, P17 has identified two clear impacts on his using Instagram to share his art that have nothing to do with the subject matter of his art and everything to do with how the platform perceives him as a “content producer.” Similarly, P22's search for inspiration produced the idea that nudes were art that people bought, but did not warn her that artistic nudes are a common target of algorithmic censorship \cite{riccio2024exposed}. Searching for inspiration is limited by platform infrastructures because it must take place within the parameters of the existing platform infrastructure, and any inspirational object found must abide by that platform's rules, and sometimes, as in the case of artistic nudes, cannot be actively searched for. \par

\subsubsection{Receptive Waiting}

    While the search for inspiration always requires a degree of mindfulness and openness to the spontaneous moments on inspiration \cite{hymer1990inspiration}, receptive waiting is characterized by the control that one can still exercise in order to capture that spontaneous moment where search produces an inspirational \textit{something} and transforms it into art. Some of the artists we spoke to took a more malleable and receptive approach to their use of search for inspiration--never drawing on one particular object or piece of media, but rather, using search to tug multiple strands of the creative outputs of others together to create their own art. P8, a 32 year old wood intarsia\footnote{A term that describes knitting together different colors and stitches to create a larger pattern or design used in woodworking.} describes how he routinely uses search to find elements of what he wants to incorporate into his woodworking:  
 
    \begin{quote}“I'm a terrible artist with a pencil. So I start online, and it's usually a Google image search and find several pictures that have elements--way ways mountains are shaped different styles of just drawing a tree [and] shit like that, that I kind of throw all in one folder, and then go back and spend a few hours in Photoshop and pull elements from this picture in this picture in this picture and put together what I actually want to make.”	\end{quote}

    P8 is not relying on any one image to find inspiration for his art, but rather is exercising control over multiple inspirational objects he discovers through search by drawing them together into a new, transformed, whole. Receptive waiting allows P8 to amass a file of images-- \textit{inspirational somethings} found through the infrastructural elements that allow for searchability of images broadly--that he then routinely edits together using Photoshop. This practice affords P8 more creative flexibility than he would if he were drawing or sketching his ideas by hand to produce an eventual inspirational object that he reproduces in wood. Search is a key element of this creative routine around inspiration where participants integrate platform infrastructures, as well as the trace elements of other creative people that are encountered through search, while not directly social, do involve the interaction between creative people and the inspiring, creative work of others.  \par

\subsection{Summary}
    Given that inspiration is different for each person, understanding the contradictory nature of inspiration is a matter of routine practice that relies on infrastructures - human, technical, and physical - for support provides a clear insight into how the creative objects and spaces where inspiration emerges are co-constitutive of that inspiration. These spaces and objects together provide the potential for inspiration to emerge and transform and be made anew once once again. \par

    In reflecting on the contradictions discussed in the previous three sections, becoming inspired is both an active and passive experience for the artists we spoke to. It is active in that we must be mindful in how we go about our everyday lives, in how we look for inspirational objects, and how we must be disciplined in how we do creative work and rely on human infrastructures for critique and inspiration. Yet inspiration is also passive, in that while we may be engaging in routine behaviors (e.g., scrolling through Instagram or Pinterest), we do not have to be looking for any idea in particular. In these spontaneous moments we may find some thread of something that we find inspirational -- waiting for it to come to us and being ready and able to act on it when the moment of spontaneous inspiration hits us. \par

\section{Discussion: Infrastructuring the Creative Internet}
    Our results show that infrastructures can be inspirational, but, more importantly, they allow for the routine enactment of and engagement with one’s creative identity through routine encounters with necessary others. The results, in particular, show that creative practice (e.g., disciplined approaches to doing creative work) are supported by platform infrastructures in how the humans who use them have built and augmented spaces for feedback, critique, and disciplined work into online artist community spaces. These spaces are vital to the development and maintenance of creative identities, as they allow for artists to interact with other artists about art.  \par

    Inspiration can be embedded into platform infrastructures, and humans appropriate and build infrastructures for creativity and inspiration. We also discuss what cannot be found in these infrastructures, or what ends up missing as a result of these infrastructures. The visual artists we spoke to agree that online platforms allow for the potential to do art because they are spaces where their creative identities are validated and supported. This creative internet is supported by the inspirational potential of the infrastructures of multiple online platforms. 

\begin{table*}

\begin{tabular}{|l|l|l|}
\hline
\multicolumn{1}{|c|}{\textbf{Characteristic}}                             	& \textbf{Necessary Other(s)}                                      	& \textbf{How They Are Used as Inspirational Objects}                                                                                                                                                                                                                                                                                                                               	\\ \hline
\multirow{2}{*}{Discipline}                                               	& Creative Peers                                                   	& \begin{tabular}[c]{@{}l@{}}* Develop and augment online community spaces to \\ promote creative practice.\\ * Provide peer support for creative practice through \\ community features (e.g., Power Hours, critique spaces).\\ * Provide feedback to and interaction with fellow creative \\ peers that solidify creative identity.\end{tabular} \\ \cline{2-3}
                                                                          	& Mentors/Teachers                                                 	& \begin{tabular}[c]{@{}l@{}}* Develop and augment online community spaces to \\ promote creative practice through feedback and critique.\\ * Provide feedback and interact with mentees or students \\ that solidify creative identity.\end{tabular}                                                                                          	\\ \hline
Spontaneity                                                               	& Creative Peers                                                   	& \begin{tabular}[c]{@{}l@{}}* Augment existing infrastructures to support interaction \\ with fellow creative peers in ways that build community.\\ * Encourage the organic development of new ideas and \\ inspirational potentials through active engagement with \\ community creative spaces.\end{tabular}                                	\\ \hline
Mindfulness                                                              	& Self                                                             	& \begin{tabular}[c]{@{}l@{}}* Being aware of one's surroundings and open to inspiration.\\ Document inspirational objects as they are encountered.\end{tabular}                                                                                                                                                                               	\\ \hline
Mindlessness                                                              	& \begin{tabular}[c]{@{}l@{}}Recommendation \\ Algorithms\end{tabular} & \begin{tabular}[c]{@{}l@{}}* Recommend content based on prior browsing history, \\ measurable types and other trace data elements.\\ * Provide diverse inspirational content.\end{tabular}                                                                                                                                                   	\\ \hline
\multirow{2}{*}{\begin{tabular}[c]{@{}l@{}}Active \\ Searching\end{tabular}}  & Search Algorithms                                                	& \begin{tabular}[c]{@{}l@{}}* Provide search results based on search terms. \\ * Index inspirational objects in findable ways.\end{tabular}                                                                                                                                                                                                   	\\ \cline{2-3}
                                                                          	& Creative Peers                                                   	& \begin{tabular}[c]{@{}l@{}}* Create and upload inspirational objects to various \\ online spaces using searchable terms.\end{tabular}                                                                                                                                                                                                        	\\ \hline
\multirow{3}{*}{\begin{tabular}[c]{@{}l@{}}Receptive \\ Waiting\end{tabular}} & Self                                                             	& \begin{tabular}[c]{@{}l@{}}* Being aware of creative goals and inspirational needs.\\ * Breaking down desired inspiration into requisite parts.\\ * Flexibility in transforming, remixing, and augmenting \\ inspirational objects.\end{tabular}                                                                                             	\\ \cline{2-3}
                                                                          	& Search Algorithms                                                	& \begin{tabular}[c]{@{}l@{}}* Provide search results based on search terms. \\ * Index inspirational objects in findable ways.\end{tabular}                                                                                                                                                                                                   	\\ \cline{2-3}
                                                                          	& \begin{tabular}[c]{@{}l@{}}Recommendation \\ Algorithms\end{tabular} & \begin{tabular}[c]{@{}l@{}}* Recommend content based on prior browsing history, \\ measurable types and other trace data elements.\\ * Provide diverse inspirational content.\end{tabular}                                                                                                                                                   	\\ \hline
\end{tabular}
\caption{A table describing the necessary others for the contradictory nature of inspiration.}
\Description[A table describing the necessary others for Inspiration.]{A table describing the necessary others for Inspiration.}
\label{tab:my-table}

\end{table*}

\subsection{The Necessary Others of The Creative Internet}

    Playing out across these collective infrastructures for inspiration is the negotiation and realization of creative identity for these artists.  Our findings show that, to become inspired, artists must interact with a series of “others” – be it other creative work, enter into creative spaces, or other artists themselves. Table \ref{tab:my-table} demonstrates the others participants described encountering. The object could be anything, but without the assemblage of human and non-human entities, the emergent circumstances for inspiration could not exist. Emergent from this table is the fact that the inspirational objects that one must form relationships with to inspire new ideas \cite{hymer1990inspiration} have no role in the table – they merely exist as it is impossible to predict how any one object will inspire any one person \cite{rudnicki2021ideas}. Inspirational objects are mediated through the platforms where they are encountered, meaning it is not the object alone, but rather the interplay between the object, the maker, the viewer, and the platform. \par

    Table \ref{tab:my-table} identifies necessary others, many of which fall into Hymer’s \cite{hymer1990inspiration} categorizations of inspirational objects – the natural (e.g., P7’s photographs of inspirational nature), the secular (e.g., creative peers and teachers). In these cases, the self is also objectified in many of the passive contradictions of inspiration, but it is mediated by technological intervention. Recommendation and search algorithms mediate the objectification of the self as an inspirational object, and, at times, give the appearance of ‘the divine’ - a “notion of an outside object temporarily occupying the inspired’s body or soul” \cite[~p.20]{hymer1990inspiration}. Necessary others are just that, necessary – for inspiration for one’s creative work, to be inspirational to others, and to the continued development and maintenance of creative identities. It is only with others that creative identity can be realized, and, therefore, it is only through relationship to others that inspiration emerges. \par

    Embedded into the socio-cultural theory of creative identity is that the performance and expression of identity is at once a personal and social task \cite{gluaveanu2014creativity}. If an individual wants to routinely interact with others while being viewed as an 'artist,' they cannot adopt that identity simply by themselves -- rather, others must see and relate to that individual as an artist \cite{gluaveanu2014creativity}. The artists we spoke to drew on online infrastructures to routinely express their creative identities, to find validation in themselves as creative people, and to engage in the collective negotiation of their creative identities. Interaction with inspirational others is vital to the development and maintenance of an artist’s creative identity. Inspiration serves as the action before the action \cite{hoppe2022before}, and our results show that inspiration is supported by an assemblage of actors - human and non-human. \par

    While creative identity can be negotiated, realized, and maintained in spaces with unknown persons \cite{gluaveanu2014creativity}, our findings show that the best support for creative identities comes through interactions with fellow creative peers in smaller online communities. Take P10 or P3, both of whom drew heavily on close personal networks on Discord to both be spontaneously inspired (P10) and to engage in the disciplined acts of creativity that work around that one moment of inspiration (P3), what Hymer refers to as the “slower, more painstaking analytic work” that tends to bracket inspiration \cite{hymer1990inspiration}. The interpersonal relationships that form between the necessary others that help support inspiration need space to organically emerge that is not afforded by many existing platform ecosystems, which tend to make construction of identity inflexible. While these platforms allow for easier expression of one’s creative identity and the connection to others, these connections only happen if the necessary others are correctly identified. \par

    Online, the co-creation of one’s creative identity now must emerge through the relations between the artist and the requisite necessary others: the algorithm and the algorithm’s datafication of the inspirational object. Unlike previous assessments, the relationship between subject and artist cannot exist anymore, as the intervention of algorithms - be they recommender systems or search algorithms - mediates what is found and where the negotiation of creative identities can emerge in a passive, rather than active way. In Hymer’s \cite{hymer1990inspiration} discussion of inspiration, we see the self objectified by ‘the divine’ as an inexplicable entity that is almost a bolt from the blue moment of inspiration. Our results, however, show the algorithmic mediation of our creative work – and therefore creative selves – serves as a transformation of \textit{our objectification of ourselves}. \par

    The mediation of technology is not divine, but rather a human-designed tool that is particularly good at tapping into our psyches and figuring out what objects may have the potential for inspiration. Take P13, browsing Pinterest and seeing the creative work of others as mediated by an algorithmic feed - she is only seeing things that the digital version of herself, based on everything she’s liked, pinned, and saved previously, wants to see. This is a double edged-sword, as it \textit{is} inspiring, however it is also limiting. Algorithmic mediation of creative work and the discovery of it limit true spontaneous moments of inspiration that emerge from seeing so outside of the mundane norm that it becomes inspirational. This impacts the potential for spontaneity in artists to create from these encounters, and it also limits how mindless someone can be – as they are constantly having to watch their feeds (as P10 explained and prior research shows \cite{simpson2022tame}) for fear of unwanted content. Further, algorithmic mediation impacts receptive waiting as there are less chances for certain inspirational objects to be discovered. Any aspect of the contradictions of inspiration that was a largely passive routine practice in an online setting (see Table\ref{tab:my-table}) is increasingly subject to the redefining of creative work in ways that are out of the control of the individual artist and are reduced to the creative intervention of how an algorithm indexes or datafies a creative object. \par

    \subsection{How Human Infrastructures Build on Existing Platform Infrastructures for Creative Support}

    Our results show that encounters with inspirational objects can be done in both active or passive ways. Table \ref{tab:my-table} shows how these interactions with inspirational others can emerge in both active and passive ways. For example, many participants described routinely looking at Pinterest (P9, P13, P15, P19) or Instagram (P3, P10, P13, P17) for inspiration, while others discussed getting inspired by music, books, or films. Inspiration emerged in this context passively, through the individual artist objectifying the inspirational object. 

    Conversely, inspiration emerged through the active and disciplined way many participants drew on infrastructures embedded into human-constructed community spaces (i.e., Discord servers) to do their creative work and then have an opportunity to share it with others for critique and feedback. Actively seeking out the creative work of others, or purposefully engaging in a creative space meant to facilitate dialogue about creative work, allows the artist to engage with the necessary others that both inspire and allow for the necessary relations that allow the artist to negotiate their creative identity. Online, creative spaces provide the infrastructure to facilitate these encounters, which both support creative identities but also foster inspiration. 

    Our findings show that the best and most effective platform infrastructures for inspiration are the spaces that are built on top of the existing platform infrastructures for creativity. These communities support disciplined creative practice - through critique from peers and mentors or teachers, they also facilitate the spontaneous moments of inspiration that emerge when creative people come together to their creative identities. These necessary others require reciprocity and community engagement from the artist.  Additionally, these human infrastructures also support receptive waiting, openness to the possibility of finding parts of a larger creative idea, as these human infrastructures facilitate the joining of otherwise unrelated ideas that inspire through communal dialogue and collective negotiation of creativity. 

    Artists look to spaces where there is a supportive and collaborative community to have meaningful interactions with other artists that not only reinforced their identity as artists, but also allowed for the organic emergence of further inspiration through the routine social interaction they support. Our participants describe drawing on online infrastructures to reach out to people on various critique-specific spaces such as Discord. These people support the routine development and maintenance of the creative identities of the artists therein \cite{gluaveanu2014creativity}. The human infrastructures promote and sustain the deep sense of security artists have in their creative selves that allows for artists routinely draw on these communities to continually negotiate their creative identities. 

    The contradictions of inspiration become muddled when non-human entities are introduced into the assemblages that help support inspiration and, in turn, creative identity formation. Many of our participants discussed their routine use of online platforms with content recommendation algorithms. These content recommendation algorithms are challenging, as they push creatives \cite{simpson2023rethinking,ma2021advertiser} and marginalized people \cite{karizat2021algorithmic,ungless2024experiences} into specific niches that flatten individual identities within broader marginalized groups \cite{lutz2024we, foryouforyou}. These platforms produce a digital version of the individual based their use of the platform and then recommend content to the measurable attributes -- or types -- of that digital version \cite{cheney2018wearedata}. Everything our participants encountered when scrolling through these feeds was mediated by how the algorithm perceived them and their art.

    This introduces a challenge for artists, as looking at people's art and becoming inspired to create art is not a new concept, nor is it one that is unique to the internet \cite{okada2017imitation}. Yet online, it the process is mediated by what the platform allows in terms of creative expression. Many participants expressed frustration with platforms like Instagram that limited what they could share. Others disliked that these platforms did not allow for purely mindless engagement, a finding that echoes the careful way that users of algorithmically curated platforms like TikTok, Pinterest, or Instagram must routinely and mindfully engage with content they encounter there \cite{simpson2022tame}. Search presents similar challenges – and while search engines are good at connecting individual artists to inspirational objects, they are not as good at connecting individuals to other humans that allow them to collectively negotiate their creative identities. 

    For example, recall how P10 encountered art on Instagram made by a person whose religious beliefs invalidated her queer identity. P10 was not able to \textit{count on} Instagram to not show her unwanted content while she casually browsed. Prior work \cite{simpson2022tame} has identified that the sensitivity of content recommendation algorithms means that one can never fully integrate use of these platforms into one's everyday routine. Our results show that, similarly, one can never just mindlessly browse for inspiration hoping for a serendipitous encounter -- there is too much chance for something to go wrong. Therefore, while online platforms support the inspiration through their ability to introduce individuals to the necessary others needed for inspiration through search or content recommendation, they are not supporting the creative identities of individual artists.

\subsection{How Platforms Can Better Support the Human Infrastructures for Creativity that Support Artist Creative Identity}

    Many of the emergent spaces that artists repeatedly pointed to as being both inspirational, and, by extension, supportive of their creative identities were community spaces built on top of existing platform infrastructures by their creative peers. Many of these places were small - curated through interpersonal connections. Platforms like Twitch, Reddit, and Discord play host to thousands of such communities, but increasing platform interest in monetization and pressure placed on artists by platforms start behaving like influencers \cite{poell2021platforms, bishop2023influencer, simpson2023rethinking} leaves many of these communities behind in terms of platform support or consideration when platform policy decisions are made. Artists that were once able to find creative peers that allowed them to negotiate their creative identities within these spaces now must look harder, and more deliberately, for the necessary human others to facilitate these moments of inspiration that help to foster, and inform creative identity. \par

    \begin{itemize}
        \item {\textbf{Design Recommendation: Fostering Human Connections For Artists}}---Platforms should move toward finding ways to put communities of people together beyond the chance encounters of their creative work facilitated by recommendation algorithm. This means that platforms need to rethink how creative work and digital identities are embodied in data, to avoid the nichification of artists. In spaces where search and recommendation algorithms play an increasing role in not only the emergence of inspiration (i.e., through mindlessness or receptive waiting), but also in the legitimization of creative identities by one's creative peers, effort should be made to diversify \textit{how} a person's art is mediated by the algorithm in a way that fosters connection between artists that are multi-faceted and diverse. Being able to support and artist's inspiration by and engagement with multiple subjects, topics, or indexable art (i.e., someone who draws comics might also do classical figure drawing) without siloing an artist into a particular niche will help artists to legitimate their creative identities in ways that are not processed through platform definitions of "success" through metrified engagement \cite{poell2021platforms,simpson2023rethinking}.  
    \end{itemize}

    While the design of community spaces for creative people are supported by platform infrastructures to some extent, they are not supported to the point where artists are flourishing on them. For example, on Discord, if one wants to upload any file over a certain size, or send a message over a certain number of characters, one must pay for Discord’s paid subscription service, Nitro. This limits how artists can engage in some of these community spaces without having to draw on other infrastructures to share and promote their creative work as they intended for it to be seen. Simpson and colleagues \cite{simpson2023captions} identify how sharing a creative product on a single platform can require an assemblage of creative tools and infrastructures, and our findings further demonstrate that having to rely on multiple infrastructures and tools to share creative work can introduce challenges for both the inspirational potential of these community spaces, but also the collective negotiation of creative identity within them.

    \begin{itemize}
        \item {\textbf{Design Recommendation: A La Carte Features for Artists}}---Platforms could move to support artist communities by developing tiers of usage for paid features that artists might find useful in fostering their creative identities – such as being able to pay to upload larger files or having various tiers of control over how a particular art object is datafied to ensure it gets to the right audience. This more menu-like approach, if extended to whole communities of artists (e.g., on a particular Discord server), rather than done on an individual level, may also create sustainability and stability of these spaces as they no longer present a financial barrier to entry for individual artists who cannot afford the subscription model. It further supports the dialectical relationship between inspiration and creative identity. In moving away from individual subscriptions to a more collective model, platforms will create more opportunities for inspiration through broader participation, and thus will provide the necessary security artists need in their construction of creative identities, to therefore produce more platform-sustaining content.
    \end{itemize}
    
    Prior work has argued that the best way to improve platforms for creatives is to decouple platform metrics from creative success\cite{simpson2023rethinking}, and other studies have pointed to the hegemonic impacts of being a "content creator" on a platform like Instagram have pushed artist toward more influencer oriented goals \cite{bishop2023influencer}. We echo this finding and encourage platforms to recognize and embrace their role as mediators for how creative identity is developed on their platforms. As such we make a final recommendation: 
    
     \begin{itemize}
        \item {\textbf{Design Recommendation: Design Policy and Features with Artists in Mind}}---Platforms should rethink their policies around common subjects in art, such as nude or semi-nude bodies, which are a foundational part of artist education and training as they are studies in an "ideal form" \cite{clark2023nude}. While this presents an unfortunate content moderation problem, platforms could consider creating artist-tailored accounts where certain content flags (e.g., nudity) are given either a) more permissive moderation or b) are moderated on a case-by-case basis. Further, platforms could develop a  more streamlined appeals process for take downs and a regionally-localized team of moderators to make these judgment calls could potentially help artists to feel better supported by platforms and foster the development of creative identities broadly. 
\end{itemize}

    While these are just suggestions, they are informed by how platform infrastructures are already being repurposed to foster and support creative identity and inspiration by individual users or communities of users. Our findings demonstrate how, online, platform infrastructures are increasingly playing a mediating role in how inspiration emerges, which necessitates the reexamination by platforms of what role they want to play in fostering creativity and creative drive in their users. Given that platforms are dependent on user-generated content to sustain their business model \cite{poell2021platforms}, considering this will be helpful as the overall platform experience on many of these online platforms progressively getting worse for many users. Because of this concerning trend, we urge platforms to consider our suggestions as a means by which to continue to support the vibrant and flourishing artistic communities that call them home.

\section{Conclusion: How Platforms Can Better Support the Human Infrastructures for Creativity that Support Artist Creative Identity} 

    This paper explored how creative identity is formed and legitimated in artists through an exploration of the contradictions of inspiration. We explored how small, niche communities of creative peers can support disciplined creative practice can facilitate the confluence of creative people and ideas such that spontaneous moments of inspiration can emerge, and creative identity can be legitimated for these artists through the routine sharing of creative work. We discussed the increasing role of algorithmic content recommendation on the contradiction of mindfulness and mindlessness - here focusing on how unstructured browsing leads to algorithmically-mediated encounters with inspirational objects, and noting that while this was a good thing, it was not without its risks as algorithmic content recommendation of potential inspirational objects was difficult to predict for many of the artists in this cohort and could potentially lead to encounters with harmful content. Finally, we explored a specific application of search on the contradiction of active searching and receptive waiting, where platform policy often presented challenges to actively searching for inspirational objects for common artistic subjects (e.g., nudes), and where control over search allowed for transformation of multiple found inspirational objects into a single creative object.

    Our findings contribute to the ongoing conversation about the mediating role that algorithms and other platform infrastructural elements play in the creative routines and development of creative identities in artists. While many of these conversations are focused on concerns over Generative AI and copyright of creative work shared online that was added to training datasets, our work focuses on the algorithms that artists are already contending with. Our findings show that while interacting with these algorithms is at times challenging, they are also playing an important role in the inspirational process, mediating passive inspirational processes such as mindlessness or receptive waiting and allowing artists to encounter a wider range of potentially inspirational objects. This represents an augmentation of the routine enactment of creative identity and work involved in searching for inspiration for artists. Inspiration and creative identity are intrinsically linked. Inspiration functions as the action that comes before the creative act, but it is also something that helps to co-construct creative identities, as each conversation around an inspirational object must come from some other inspirational object and the artist who found it to be inspiring.
\begin{acks}
We'd like to thank our participants who shared their expertise with us. Without them, this research could not exist. We'd also like to thank, Mona Sloane, Morgan Klaus Scheuerman, and Jordan Taylor, for their thoughtful feedback on early drafts. We'd also like to thank the reviewers, who provided robust feedback that helped make this paper what it is today.
\end{acks}

\bibliographystyle{ACM-Reference-Format}
\bibliography{References}


\begin{thebibliography}{79}


\ifx \showCODEN    \undefined \def \showCODEN     #1{\unskip}     \fi
\ifx \showISBNx    \undefined \def \showISBNx     #1{\unskip}     \fi
\ifx \showISBNxiii \undefined \def \showISBNxiii  #1{\unskip}     \fi
\ifx \showISSN     \undefined \def \showISSN      #1{\unskip}     \fi
\ifx \showLCCN     \undefined \def \showLCCN      #1{\unskip}     \fi
\ifx \shownote     \undefined \def \shownote      #1{#1}          \fi
\ifx \showarticletitle \undefined \def \showarticletitle #1{#1}   \fi
\ifx \showURL      \undefined \def \showURL       {\relax}        \fi
\providecommand\bibfield[2]{#2}
\providecommand\bibinfo[2]{#2}
\providecommand\natexlab[1]{#1}
\providecommand\showeprint[2][]{arXiv:#2}

\bibitem[Amabile and Conti(1999)]%
        {amabile1999changes}
\bibfield{author}{\bibinfo{person}{Teresa~M Amabile} {and} \bibinfo{person}{Regina Conti}.} \bibinfo{year}{1999}\natexlab{}.
\newblock \showarticletitle{Changes in the work environment for creativity during downsizing}.
\newblock \bibinfo{journal}{\emph{Academy of Management journal}} \bibinfo{volume}{42}, \bibinfo{number}{6} (\bibinfo{year}{1999}), \bibinfo{pages}{630--640}.
\newblock


\bibitem[Anderson and Vyas(2022)]%
        {AndersonShredding2022}
\bibfield{author}{\bibinfo{person}{India Anderson} {and} \bibinfo{person}{Dhaval Vyas}.} \bibinfo{year}{2022}\natexlab{}.
\newblock \showarticletitle{Shedding Ageist Perceptions of Making: Creativity in Older Adult Maker Communities}. In \bibinfo{booktitle}{\emph{Proceedings of the 14th Conference on Creativity and Cognition}} (Venice, Italy) \emph{(\bibinfo{series}{C\&C '22})}. \bibinfo{publisher}{Association for Computing Machinery}, \bibinfo{address}{New York, NY, USA}, \bibinfo{pages}{208–219}.
\newblock
\showISBNx{9781450393270}
\href{https://doi.org/10.1145/3527927.3532800}{doi:\nolinkurl{10.1145/3527927.3532800}}


\bibitem[Biernacki and Waldorf(1981)]%
        {biernacki1981snowball}
\bibfield{author}{\bibinfo{person}{Patrick Biernacki} {and} \bibinfo{person}{Dan Waldorf}.} \bibinfo{year}{1981}\natexlab{}.
\newblock \showarticletitle{Snowball sampling: Problems and techniques of chain referral sampling}.
\newblock \bibinfo{journal}{\emph{Sociological Methods \& Research}} \bibinfo{volume}{10}, \bibinfo{number}{2} (\bibinfo{year}{1981}), \bibinfo{pages}{141--163}.
\newblock


\bibitem[Bishop(2019)]%
        {bishop2019managing}
\bibfield{author}{\bibinfo{person}{Sophie Bishop}.} \bibinfo{year}{2019}\natexlab{}.
\newblock \showarticletitle{Managing visibility on YouTube through algorithmic gossip}.
\newblock \bibinfo{journal}{\emph{New Media \& Society}} \bibinfo{volume}{21}, \bibinfo{number}{11-12} (\bibinfo{year}{2019}), \bibinfo{pages}{2589--2606}.
\newblock


\bibitem[Bishop(2020)]%
        {bishop2020algorithmic}
\bibfield{author}{\bibinfo{person}{Sophie Bishop}.} \bibinfo{year}{2020}\natexlab{}.
\newblock \showarticletitle{Algorithmic experts: Selling algorithmic lore on YouTube}.
\newblock \bibinfo{journal}{\emph{Social Media+ Society}} \bibinfo{volume}{6}, \bibinfo{number}{1} (\bibinfo{year}{2020}), \bibinfo{pages}{2056305119897323}.
\newblock


\bibitem[Bishop(2023)]%
        {bishop2023influencer}
\bibfield{author}{\bibinfo{person}{Sophie Bishop}.} \bibinfo{year}{2023}\natexlab{}.
\newblock \showarticletitle{Influencer creep: How artists strategically navigate the platformisation of art worlds}.
\newblock \bibinfo{journal}{\emph{New media \& society}} (\bibinfo{year}{2023}), \bibinfo{pages}{14614448231206090}.
\newblock


\bibitem[Bowker(1994)]%
        {bowker1994information}
\bibfield{author}{\bibinfo{person}{Geoffrey Bowker}.} \bibinfo{year}{1994}\natexlab{}.
\newblock \showarticletitle{Information mythology: The world of/as information}.
\newblock \bibinfo{journal}{\emph{Information acumen: The understanding and use of knowledge in modern business}} (\bibinfo{year}{1994}), \bibinfo{pages}{231--247}.
\newblock


\bibitem[Bowker and Star(1999)]%
        {bowker2000sorting}
\bibfield{author}{\bibinfo{person}{Geoffrey~C Bowker} {and} \bibinfo{person}{Susan~Leigh Star}.} \bibinfo{year}{1999}\natexlab{}.
\newblock \bibinfo{booktitle}{\emph{Sorting things out: Classification and its consequences}}.
\newblock \bibinfo{publisher}{MIT press}.
\newblock


\bibitem[Britton et~al\mbox{.}(2019)]%
        {britton2019mothers}
\bibfield{author}{\bibinfo{person}{Lauren Britton}, \bibinfo{person}{Louise Barkhuus}, {and} \bibinfo{person}{Bryan Semaan}.} \bibinfo{year}{2019}\natexlab{}.
\newblock \showarticletitle{"Mothers as Candy Wrappers": Critical Infrastructure Supporting the Transition into Motherhood}.
\newblock \bibinfo{journal}{\emph{Proc. ACM Hum.-Comput. Interact.}} \bibinfo{volume}{3}, \bibinfo{number}{GROUP}, Article \bibinfo{articleno}{232} (\bibinfo{date}{dec} \bibinfo{year}{2019}), \bibinfo{numpages}{21}~pages.
\newblock
\href{https://doi.org/10.1145/3361113}{doi:\nolinkurl{10.1145/3361113}}


\bibitem[Cheney-Lippold(2018)]%
        {cheney2018wearedata}
\bibfield{author}{\bibinfo{person}{John Cheney-Lippold}.} \bibinfo{year}{2018}\natexlab{}.
\newblock \bibinfo{booktitle}{\emph{We are data: Algorithms and the making of our digital selves}}.
\newblock \bibinfo{publisher}{NYU Press}.
\newblock


\bibitem[Cirillo(2018)]%
        {cirillo2018pomodoro}
\bibfield{author}{\bibinfo{person}{Francesco Cirillo}.} \bibinfo{year}{2018}\natexlab{}.
\newblock \bibinfo{booktitle}{\emph{The Pomodoro technique: The acclaimed time-management system that has transformed how we work}}.
\newblock \bibinfo{publisher}{Currency}.
\newblock


\bibitem[Clark(2023)]%
        {clark2023nude}
\bibfield{author}{\bibinfo{person}{Kenneth Clark}.} \bibinfo{year}{2023}\natexlab{}.
\newblock \bibinfo{booktitle}{\emph{The nude: A study in ideal form}}. Vol.~\bibinfo{volume}{2}.
\newblock \bibinfo{publisher}{Princeton University Press}.
\newblock


\bibitem[Cook and Brown(1999)]%
        {cook1999bridging}
\bibfield{author}{\bibinfo{person}{Scott~DN Cook} {and} \bibinfo{person}{John~Seely Brown}.} \bibinfo{year}{1999}\natexlab{}.
\newblock \showarticletitle{Bridging epistemologies: The generative dance between organizational knowledge and organizational knowing}.
\newblock \bibinfo{journal}{\emph{Organization science}} \bibinfo{volume}{10}, \bibinfo{number}{4} (\bibinfo{year}{1999}), \bibinfo{pages}{381--400}.
\newblock


\bibitem[De~Rond(2014)]%
        {derond2014structure}
\bibfield{author}{\bibinfo{person}{Mark De~Rond}.} \bibinfo{year}{2014}\natexlab{}.
\newblock \showarticletitle{The structure of serendipity}.
\newblock \bibinfo{journal}{\emph{Culture and Organization}} \bibinfo{volume}{20}, \bibinfo{number}{5} (\bibinfo{year}{2014}), \bibinfo{pages}{342--358}.
\newblock


\bibitem[Desai(2024)]%
        {desai2024psychology}
\bibfield{author}{\bibinfo{person}{Maya Desai}.} \bibinfo{year}{2024}\natexlab{}.
\newblock \showarticletitle{The Psychology of Inspiration: Its Impact on Creative Processes}.
\newblock \bibinfo{journal}{\emph{Shodh Sagar Journal of Inspiration and Psychology}} \bibinfo{volume}{1}, \bibinfo{number}{2} (\bibinfo{year}{2024}), \bibinfo{pages}{36--41}.
\newblock


\bibitem[DeVito et~al\mbox{.}(2018)]%
        {devito2018too}
\bibfield{author}{\bibinfo{person}{Michael~Ann DeVito}, \bibinfo{person}{Ashley~Marie Walker}, {and} \bibinfo{person}{Jeremy Birnholtz}.} \bibinfo{year}{2018}\natexlab{}.
\newblock \showarticletitle{'Too Gay for Facebook' Presenting LGBTQ+ Identity Throughout the Personal Social Media Ecosystem}.
\newblock \bibinfo{journal}{\emph{Proceedings of the ACM on Human-Computer Interaction}} \bibinfo{volume}{2}, \bibinfo{number}{CSCW} (\bibinfo{year}{2018}), \bibinfo{pages}{1--23}.
\newblock


\bibitem[Dimond et~al\mbox{.}(2012)]%
        {dimond2012qualitative}
\bibfield{author}{\bibinfo{person}{Jill~P Dimond}, \bibinfo{person}{Casey Fiesler}, \bibinfo{person}{Betsy DiSalvo}, \bibinfo{person}{Jon Pelc}, {and} \bibinfo{person}{Amy~S Bruckman}.} \bibinfo{year}{2012}\natexlab{}.
\newblock \showarticletitle{Qualitative data collection technologies: A comparison of instant messaging, email, and phone}. In \bibinfo{booktitle}{\emph{Proceedings of the 2012 ACM International Conference on Supporting Group Work}}. \bibinfo{pages}{277--280}.
\newblock


\bibitem[Duffy et~al\mbox{.}(2021)]%
        {duffy2021nested}
\bibfield{author}{\bibinfo{person}{Brooke~Erin Duffy}, \bibinfo{person}{Annika Pinch}, \bibinfo{person}{Shruti Sannon}, {and} \bibinfo{person}{Megan Sawey}.} \bibinfo{year}{2021}\natexlab{}.
\newblock \showarticletitle{The Nested Precarities of Creative Labor on Social Media}.
\newblock \bibinfo{journal}{\emph{Social Media+ Society}} \bibinfo{volume}{7}, \bibinfo{number}{2} (\bibinfo{year}{2021}).
\newblock


\bibitem[Edwards(2003)]%
        {edwards2003infrastructure}
\bibfield{author}{\bibinfo{person}{Paul~N Edwards}.} \bibinfo{year}{2003}\natexlab{}.
\newblock \showarticletitle{Infrastructure and modernity: Force, time, and social organization in the history of sociotechnical systems}.
\newblock \bibinfo{journal}{\emph{Modernity and technology}}  \bibinfo{volume}{1} (\bibinfo{year}{2003}), \bibinfo{pages}{185--226}.
\newblock


\bibitem[Emerson et~al\mbox{.}(2024)]%
        {EmersonShared2024}
\bibfield{author}{\bibinfo{person}{Adam~G. Emerson}, \bibinfo{person}{Shreyosi Endow}, {and} \bibinfo{person}{Cesar Torres}.} \bibinfo{year}{2024}\natexlab{}.
\newblock \showarticletitle{Shared, Shaped, and Stolen: Tracing Sites of Knowledge Transfer across Creative Communities of Practice}. In \bibinfo{booktitle}{\emph{Proceedings of the 16th Conference on Creativity \& Cognition}} (Chicago, IL, USA) \emph{(\bibinfo{series}{C\&C '24})}. \bibinfo{publisher}{Association for Computing Machinery}, \bibinfo{address}{New York, NY, USA}, \bibinfo{pages}{638–650}.
\newblock
\showISBNx{9798400704857}
\href{https://doi.org/10.1145/3635636.3656199}{doi:\nolinkurl{10.1145/3635636.3656199}}


\bibitem[Feldman(2000)]%
        {feldman2000organizational}
\bibfield{author}{\bibinfo{person}{Martha~S Feldman}.} \bibinfo{year}{2000}\natexlab{}.
\newblock \showarticletitle{Organizational routines as a source of continuous change}.
\newblock \bibinfo{journal}{\emph{Organization science}} \bibinfo{volume}{11}, \bibinfo{number}{6} (\bibinfo{year}{2000}), \bibinfo{pages}{611--629}.
\newblock


\bibitem[Feldman et~al\mbox{.}(2016)]%
        {feldman2016beyond}
\bibfield{author}{\bibinfo{person}{Martha~S Feldman}, \bibinfo{person}{Brian~T Pentland}, \bibinfo{person}{Luciana D’Adderio}, {and} \bibinfo{person}{Nathalie Lazaric}.} \bibinfo{year}{2016}\natexlab{}.
\newblock \bibinfo{title}{Beyond routines as things: Introduction to the special issue on routine dynamics}.
\newblock \bibinfo{numpages}{505--513}~pages.
\newblock


\bibitem[Foster and Ford(2003)]%
        {foster2003serendipity}
\bibfield{author}{\bibinfo{person}{Allen Foster} {and} \bibinfo{person}{Nigel Ford}.} \bibinfo{year}{2003}\natexlab{}.
\newblock \showarticletitle{Serendipity and information seeking: an empirical study}.
\newblock \bibinfo{journal}{\emph{Journal of documentation}} \bibinfo{volume}{59}, \bibinfo{number}{3} (\bibinfo{year}{2003}), \bibinfo{pages}{321--340}.
\newblock


\bibitem[Frich et~al\mbox{.}(2019)]%
        {frich2019mapping}
\bibfield{author}{\bibinfo{person}{Jonas Frich}, \bibinfo{person}{Lindsay MacDonald~Vermeulen}, \bibinfo{person}{Christian Remy}, \bibinfo{person}{Michael~Mose Biskjaer}, {and} \bibinfo{person}{Peter Dalsgaard}.} \bibinfo{year}{2019}\natexlab{}.
\newblock \showarticletitle{Mapping the landscape of creativity support tools in HCI}. In \bibinfo{booktitle}{\emph{Proceedings of the 2019 CHI Conference on Human Factors in Computing Systems}}. \bibinfo{pages}{1--18}.
\newblock


\bibitem[Frich et~al\mbox{.}(2018)]%
        {frich2018twenty}
\bibfield{author}{\bibinfo{person}{Jonas Frich}, \bibinfo{person}{Michael Mose~Biskjaer}, {and} \bibinfo{person}{Peter Dalsgaard}.} \bibinfo{year}{2018}\natexlab{}.
\newblock \showarticletitle{Twenty years of creativity research in human-computer interaction: Current state and future directions}. In \bibinfo{booktitle}{\emph{Proceedings of the 2018 Designing Interactive Systems Conference}}. \bibinfo{pages}{1235--1257}.
\newblock


\bibitem[Friske et~al\mbox{.}(2020)]%
        {friske2020entangling}
\bibfield{author}{\bibinfo{person}{Mikhaila Friske}, \bibinfo{person}{Jordan Wirfs-Brock}, {and} \bibinfo{person}{Laura Devendorf}.} \bibinfo{year}{2020}\natexlab{}.
\newblock \showarticletitle{Entangling the roles of maker and interpreter in interpersonal data narratives: Explorations in yarn and sound}. In \bibinfo{booktitle}{\emph{Proceedings of the 2020 ACM Designing Interactive Systems Conference}}. \bibinfo{pages}{297--310}.
\newblock


\bibitem[Gallagher(2017)]%
        {GallagherIdeation2017}
\bibfield{author}{\bibinfo{person}{Courtney~Lynn Gallagher}.} \bibinfo{year}{2017}\natexlab{}.
\newblock \showarticletitle{Sketching for Ideation: A Structured Approach for Increasing Divergent Thinking}. In \bibinfo{booktitle}{\emph{Proceedings of the 2017 CHI Conference Extended Abstracts on Human Factors in Computing Systems}} (Denver, Colorado, USA) \emph{(\bibinfo{series}{CHI EA '17})}. \bibinfo{publisher}{Association for Computing Machinery}, \bibinfo{address}{New York, NY, USA}, \bibinfo{pages}{106–111}.
\newblock
\showISBNx{9781450346566}
\href{https://doi.org/10.1145/3027063.3048424}{doi:\nolinkurl{10.1145/3027063.3048424}}


\bibitem[Gallay(2013)]%
        {gallay2013understanding}
\bibfield{author}{\bibinfo{person}{Lillian~Hemingway Gallay}.} \bibinfo{year}{2013}\natexlab{}.
\newblock \bibinfo{booktitle}{\emph{Understanding and treating creative block in professional artists}}.
\newblock \bibinfo{publisher}{Alliant International University}.
\newblock


\bibitem[Gecas(1982)]%
        {gecas1982self}
\bibfield{author}{\bibinfo{person}{Viktor Gecas}.} \bibinfo{year}{1982}\natexlab{}.
\newblock \showarticletitle{The self-concept}.
\newblock \bibinfo{journal}{\emph{Annual Review of Sociology}} \bibinfo{volume}{8}, \bibinfo{number}{1} (\bibinfo{year}{1982}), \bibinfo{pages}{1--33}.
\newblock


\bibitem[Giddens(1991)]%
        {giddens1991modernity}
\bibfield{author}{\bibinfo{person}{Anthony Giddens}.} \bibinfo{year}{1991}\natexlab{}.
\newblock \bibinfo{booktitle}{\emph{Modernity and self-identity: Self and society in the late modern age}}.
\newblock \bibinfo{publisher}{Stanford university press}.
\newblock


\bibitem[Gl{\u{a}}veanu and Tanggaard(2014)]%
        {gluaveanu2014creativity}
\bibfield{author}{\bibinfo{person}{Vlad~Petre Gl{\u{a}}veanu} {and} \bibinfo{person}{Lene Tanggaard}.} \bibinfo{year}{2014}\natexlab{}.
\newblock \showarticletitle{Creativity, identity, and representation: Towards a socio-cultural theory of creative identity}.
\newblock \bibinfo{journal}{\emph{New Ideas in Psychology}}  \bibinfo{volume}{34} (\bibinfo{year}{2014}), \bibinfo{pages}{12--21}.
\newblock


\bibitem[Hanseth and Lyytinen(2010)]%
        {hanseth2010design}
\bibfield{author}{\bibinfo{person}{Ole Hanseth} {and} \bibinfo{person}{Kalle Lyytinen}.} \bibinfo{year}{2010}\natexlab{}.
\newblock \showarticletitle{Design theory for dynamic complexity in information infrastructures: the case of building internet}.
\newblock \bibinfo{journal}{\emph{Journal of Information Technology}} \bibinfo{volume}{25}, \bibinfo{number}{1} (\bibinfo{year}{2010}), \bibinfo{pages}{1--19}.
\newblock


\bibitem[Hart(1998)]%
        {hart1998inspiration}
\bibfield{author}{\bibinfo{person}{Tobin Hart}.} \bibinfo{year}{1998}\natexlab{}.
\newblock \showarticletitle{Inspiration: Exploring the experience and its meaning}.
\newblock \bibinfo{journal}{\emph{Journal of Humanistic Psychology}} \bibinfo{volume}{38}, \bibinfo{number}{3} (\bibinfo{year}{1998}), \bibinfo{pages}{7--35}.
\newblock


\bibitem[Hill et~al\mbox{.}(2016)]%
        {hill2016searching}
\bibfield{author}{\bibinfo{person}{Timothy Hill}, \bibinfo{person}{Valentine Charles}, \bibinfo{person}{Juliane Stiller}, {and} \bibinfo{person}{Antoine Isaac}.} \bibinfo{year}{2016}\natexlab{}.
\newblock \showarticletitle{“Searching for inspiration”: User needs and search architecture in Europeana collections}.
\newblock \bibinfo{journal}{\emph{Proceedings of the Association for Information Science and Technology}} \bibinfo{volume}{53}, \bibinfo{number}{1} (\bibinfo{year}{2016}), \bibinfo{pages}{1--7}.
\newblock


\bibitem[Hoppe(2022)]%
        {hoppe2022before}
\bibfield{author}{\bibinfo{person}{Alexander~D Hoppe}.} \bibinfo{year}{2022}\natexlab{}.
\newblock \showarticletitle{Before creativity: Inspiration as a micro foundation for action}. In \bibinfo{booktitle}{\emph{Sociological Forum}}, Vol.~\bibinfo{volume}{37}. Wiley Online Library, \bibinfo{pages}{350--368}.
\newblock


\bibitem[Houston et~al\mbox{.}(2016)]%
        {JacksonValuesInrepair}
\bibfield{author}{\bibinfo{person}{Lara Houston}, \bibinfo{person}{Steven~J. Jackson}, \bibinfo{person}{Daniela~K. Rosner}, \bibinfo{person}{Syed~Ishtiaque Ahmed}, \bibinfo{person}{Meg Young}, {and} \bibinfo{person}{Laewoo Kang}.} \bibinfo{year}{2016}\natexlab{}.
\newblock \showarticletitle{Values in Repair}. In \bibinfo{booktitle}{\emph{Proceedings of the 2016 CHI Conference on Human Factors in Computing Systems}} (San Jose, California, USA) \emph{(\bibinfo{series}{CHI '16})}. \bibinfo{publisher}{Association for Computing Machinery}, \bibinfo{address}{New York, NY, USA}, \bibinfo{pages}{1403–1414}.
\newblock
\showISBNx{9781450333627}
\href{https://doi.org/10.1145/2858036.2858470}{doi:\nolinkurl{10.1145/2858036.2858470}}


\bibitem[Hwang and Won(2021)]%
        {hwangideabot2021}
\bibfield{author}{\bibinfo{person}{Angel Hsing-Chi Hwang} {and} \bibinfo{person}{Andrea~Stevenson Won}.} \bibinfo{year}{2021}\natexlab{}.
\newblock \showarticletitle{IdeaBot: investigating social facilitation in human-machine team creativity}. In \bibinfo{booktitle}{\emph{Proceedings of the 2021 CHI conference on human factors in computing systems}}. \bibinfo{pages}{1--16}.
\newblock


\bibitem[Hymer(1990)]%
        {hymer1990inspiration}
\bibfield{author}{\bibinfo{person}{Sharon Hymer}.} \bibinfo{year}{1990}\natexlab{}.
\newblock \showarticletitle{On inspiration}.
\newblock \bibinfo{journal}{\emph{The Psychotherapy Patient}} \bibinfo{volume}{6}, \bibinfo{number}{3-4} (\bibinfo{year}{1990}), \bibinfo{pages}{17--38}.
\newblock


\bibitem[Ibarra and Petriglieri(2010)]%
        {ibarra2010identity}
\bibfield{author}{\bibinfo{person}{Herminia Ibarra} {and} \bibinfo{person}{Jennifer~L Petriglieri}.} \bibinfo{year}{2010}\natexlab{}.
\newblock \showarticletitle{Identity work and play.}
\newblock \bibinfo{journal}{\emph{Journal of Organizational Change Management}} (\bibinfo{year}{2010}).
\newblock


\bibitem[Jones et~al\mbox{.}(2024)]%
        {JonesHandSpinning2024}
\bibfield{author}{\bibinfo{person}{Lee Jones}, \bibinfo{person}{Ahmed Awad}, \bibinfo{person}{Marion Koelle}, {and} \bibinfo{person}{Sara Nabil}.} \bibinfo{year}{2024}\natexlab{}.
\newblock \showarticletitle{Hand Spinning E-textile Yarns: Understanding the Craft Practices of Hand Spinners and Workshop Explorations with E-textile Fibers and Materials}. In \bibinfo{booktitle}{\emph{Proceedings of the 2024 ACM Designing Interactive Systems Conference}} (Copenhagen, Denmark) \emph{(\bibinfo{series}{DIS '24})}. \bibinfo{publisher}{Association for Computing Machinery}, \bibinfo{address}{New York, NY, USA}, \bibinfo{pages}{1–19}.
\newblock
\showISBNx{9798400705830}
\href{https://doi.org/10.1145/3643834.3660717}{doi:\nolinkurl{10.1145/3643834.3660717}}


\bibitem[Karimi et~al\mbox{.}(2020)]%
        {karimi2020creative}
\bibfield{author}{\bibinfo{person}{Pegah Karimi}, \bibinfo{person}{Jeba Rezwana}, \bibinfo{person}{Safat Siddiqui}, \bibinfo{person}{Mary~Lou Maher}, {and} \bibinfo{person}{Nasrin Dehbozorgi}.} \bibinfo{year}{2020}\natexlab{}.
\newblock \showarticletitle{Creative sketching partner: an analysis of human-AI co-creativity}. In \bibinfo{booktitle}{\emph{Proceedings of the 25th international conference on intelligent user interfaces}}. \bibinfo{pages}{221--230}.
\newblock


\bibitem[Karizat et~al\mbox{.}(2021)]%
        {karizat2021algorithmic}
\bibfield{author}{\bibinfo{person}{Nadia Karizat}, \bibinfo{person}{Dan Delmonaco}, \bibinfo{person}{Motahhare Eslami}, {and} \bibinfo{person}{Nazanin Andalibi}.} \bibinfo{year}{2021}\natexlab{}.
\newblock \showarticletitle{Algorithmic folk theories and identity: How TikTok users co-produce Knowledge of identity and engage in algorithmic resistance}.
\newblock \bibinfo{journal}{\emph{Proceedings of the ACM on Human-Computer Interaction}} \bibinfo{volume}{5}, \bibinfo{number}{CSCW2} (\bibinfo{year}{2021}), \bibinfo{pages}{1--44}.
\newblock


\bibitem[Laing and Masoodian(2015)]%
        {laing2015study}
\bibfield{author}{\bibinfo{person}{Simon Laing} {and} \bibinfo{person}{Masood Masoodian}.} \bibinfo{year}{2015}\natexlab{}.
\newblock \showarticletitle{A study of the role of visual information in supporting ideation in graphic design}.
\newblock \bibinfo{journal}{\emph{Journal of the Association for Information Science and Technology}} \bibinfo{volume}{66}, \bibinfo{number}{6} (\bibinfo{year}{2015}), \bibinfo{pages}{1199--1211}.
\newblock


\bibitem[Latour(2007)]%
        {latour2007reassembling}
\bibfield{author}{\bibinfo{person}{Bruno Latour}.} \bibinfo{year}{2007}\natexlab{}.
\newblock \bibinfo{booktitle}{\emph{Reassembling the social: An introduction to actor-network-theory}}.
\newblock \bibinfo{publisher}{Oxford University Press}.
\newblock


\bibitem[Lee et~al\mbox{.}(2006)]%
        {LeeDourishMark2006}
\bibfield{author}{\bibinfo{person}{Charlotte~P. Lee}, \bibinfo{person}{Paul Dourish}, {and} \bibinfo{person}{Gloria Mark}.} \bibinfo{year}{2006}\natexlab{}.
\newblock \showarticletitle{The Human Infrastructure of Cyberinfrastructure}. In \bibinfo{booktitle}{\emph{Proceedings of the 2006 20th Anniversary Conference on Computer Supported Cooperative Work}} (Banff, Alberta, Canada) \emph{(\bibinfo{series}{CSCW '06})}. \bibinfo{publisher}{Association for Computing Machinery}, \bibinfo{address}{New York, NY, USA}, \bibinfo{pages}{483–492}.
\newblock
\showISBNx{1595932496}
\href{https://doi.org/10.1145/1180875.1180950}{doi:\nolinkurl{10.1145/1180875.1180950}}


\bibitem[Lin et~al\mbox{.}(2020)]%
        {LinCollaborativeIdeation2020}
\bibfield{author}{\bibinfo{person}{Yuyu Lin}, \bibinfo{person}{Jiahao Guo}, \bibinfo{person}{Yang Chen}, \bibinfo{person}{Cheng Yao}, {and} \bibinfo{person}{Fangtian Ying}.} \bibinfo{year}{2020}\natexlab{}.
\newblock \showarticletitle{It Is Your Turn: Collaborative Ideation With a Co-Creative Robot through Sketch}. In \bibinfo{booktitle}{\emph{Proceedings of the 2020 CHI Conference on Human Factors in Computing Systems}} (Honolulu, HI, USA) \emph{(\bibinfo{series}{CHI '20})}. \bibinfo{publisher}{Association for Computing Machinery}, \bibinfo{address}{New York, NY, USA}, \bibinfo{pages}{1–14}.
\newblock
\showISBNx{9781450367080}
\href{https://doi.org/10.1145/3313831.3376258}{doi:\nolinkurl{10.1145/3313831.3376258}}


\bibitem[Lucero(2012)]%
        {lucero2012framing}
\bibfield{author}{\bibinfo{person}{Andr{\'e}s Lucero}.} \bibinfo{year}{2012}\natexlab{}.
\newblock \showarticletitle{Framing, aligning, paradoxing, abstracting, and directing: how design mood boards work}. In \bibinfo{booktitle}{\emph{Proceedings of the designing interactive systems conference}}. \bibinfo{pages}{438--447}.
\newblock


\bibitem[Lutz and Aragon(2024)]%
        {lutz2024we}
\bibfield{author}{\bibinfo{person}{Nina Lutz} {and} \bibinfo{person}{Cecilia Aragon}.} \bibinfo{year}{2024}\natexlab{}.
\newblock \showarticletitle{" We're not all construction workers": Algorithmic Compression of Latinidad on TikTok}.
\newblock \bibinfo{journal}{\emph{arXiv preprint arXiv:2407.13927}} (\bibinfo{year}{2024}).
\newblock


\bibitem[Ma et~al\mbox{.}(2023)]%
        {ma2023multi}
\bibfield{author}{\bibinfo{person}{Renkai Ma}, \bibinfo{person}{Xinning Gui}, {and} \bibinfo{person}{Yubo Kou}.} \bibinfo{year}{2023}\natexlab{}.
\newblock \showarticletitle{Multi-Platform Content Creation: The Configuration of Creator Ecology through Platform Prioritization, Content Synchronization, and Audience Management}. In \bibinfo{booktitle}{\emph{Proceedings of the 2023 CHI Conference on Human Factors in Computing Systems}}. \bibinfo{pages}{1--19}.
\newblock


\bibitem[Ma and Kou(2021)]%
        {ma2021advertiser}
\bibfield{author}{\bibinfo{person}{Renkai Ma} {and} \bibinfo{person}{Yubo Kou}.} \bibinfo{year}{2021}\natexlab{}.
\newblock \showarticletitle{" How advertiser-friendly is my video?": YouTuber's Socioeconomic Interactions with Algorithmic Content Moderation}.
\newblock \bibinfo{journal}{\emph{Proceedings of the ACM on Human-Computer Interaction}} \bibinfo{volume}{5}, \bibinfo{number}{CSCW2} (\bibinfo{year}{2021}), \bibinfo{pages}{1--25}.
\newblock


\bibitem[Morse(1970)]%
        {morse1970browsing}
\bibfield{author}{\bibinfo{person}{Philip~M Morse}.} \bibinfo{year}{1970}\natexlab{}.
\newblock \showarticletitle{On Browsing: The Use of Search Theory in the Search for Information.}
\newblock \bibinfo{journal}{\emph{MIT Operations Research Center}} (\bibinfo{year}{1970}).
\newblock
\urldef\tempurl%
\url{https://apps.dtic.mil/sti/tr/pdf/AD0702920.pdf}
\showURL{%
\tempurl}


\bibitem[Okada and Ishibashi(2017)]%
        {okada2017imitation}
\bibfield{author}{\bibinfo{person}{Takeshi Okada} {and} \bibinfo{person}{Kentaro Ishibashi}.} \bibinfo{year}{2017}\natexlab{}.
\newblock \showarticletitle{Imitation, inspiration, and creation: Cognitive process of creative drawing by copying others' artworks}.
\newblock \bibinfo{journal}{\emph{Cognitive science}} \bibinfo{volume}{41}, \bibinfo{number}{7} (\bibinfo{year}{2017}), \bibinfo{pages}{1804--1837}.
\newblock


\bibitem[O'Raghallaigh et~al\mbox{.}(2017)]%
        {oraghallaigh2017sociomateriality}
\bibfield{author}{\bibinfo{person}{Paidi O'Raghallaigh}, \bibinfo{person}{Stephen McCarthy}, {and} \bibinfo{person}{Fr{\'e}d{\'e}ric Adam}.} \bibinfo{year}{2017}\natexlab{}.
\newblock \showarticletitle{Sociomateriality: an object inspired proposal for IS scholars}.
\newblock \bibinfo{journal}{\emph{Proceedings of the 25th European Conference on Information Systems (ECIS)}} (\bibinfo{year}{2017}).
\newblock


\bibitem[Orlikowski(2007)]%
        {orlikowski2007sociomaterial}
\bibfield{author}{\bibinfo{person}{Wanda~J Orlikowski}.} \bibinfo{year}{2007}\natexlab{}.
\newblock \showarticletitle{Sociomaterial practices: Exploring technology at work}.
\newblock \bibinfo{journal}{\emph{Organization studies}} \bibinfo{volume}{28}, \bibinfo{number}{9} (\bibinfo{year}{2007}), \bibinfo{pages}{1435--1448}.
\newblock


\bibitem[Pentland et~al\mbox{.}(2012)]%
        {pentland2012dynamics}
\bibfield{author}{\bibinfo{person}{Brian~T Pentland}, \bibinfo{person}{Martha~S Feldman}, \bibinfo{person}{Markus~C Becker}, {and} \bibinfo{person}{Peng Liu}.} \bibinfo{year}{2012}\natexlab{}.
\newblock \showarticletitle{Dynamics of organizational routines: A generative model}.
\newblock \bibinfo{journal}{\emph{Journal of Management Studies}} \bibinfo{volume}{49}, \bibinfo{number}{8} (\bibinfo{year}{2012}), \bibinfo{pages}{1484--1508}.
\newblock


\bibitem[Perriton(2023)]%
        {perriton2023constitutive}
\bibfield{author}{\bibinfo{person}{Emma Perriton}.} \bibinfo{year}{2023}\natexlab{}.
\newblock \showarticletitle{The constitutive entanglement between open office spacing and grouping: The production of sociomaterial control}.
\newblock \bibinfo{journal}{\emph{Organization}} (\bibinfo{year}{2023}), \bibinfo{pages}{13505084231198449}.
\newblock


\bibitem[Poell et~al\mbox{.}(2021)]%
        {poell2021platforms}
\bibfield{author}{\bibinfo{person}{Thomas Poell}, \bibinfo{person}{David~B Nieborg}, {and} \bibinfo{person}{Brooke~Erin Duffy}.} \bibinfo{year}{2021}\natexlab{}.
\newblock \bibinfo{booktitle}{\emph{Platforms and cultural production}}.
\newblock \bibinfo{publisher}{John Wiley \& Sons}.
\newblock


\bibitem[Riccio et~al\mbox{.}(2024)]%
        {riccio2024exposed}
\bibfield{author}{\bibinfo{person}{Piera Riccio}, \bibinfo{person}{Thomas Hofmann}, {and} \bibinfo{person}{Nuria Oliver}.} \bibinfo{year}{2024}\natexlab{}.
\newblock \showarticletitle{Exposed or Erased: Algorithmic Censorship of Nudity in Art}. In \bibinfo{booktitle}{\emph{Proceedings of the CHI Conference on Human Factors in Computing Systems}}. \bibinfo{pages}{1--17}.
\newblock


\bibitem[Rice et~al\mbox{.}(2001)]%
        {rice2001accessing}
\bibfield{author}{\bibinfo{person}{Ronald~E Rice}, \bibinfo{person}{Maureen McCreadie}, {and} \bibinfo{person}{Shan-Ju~L Chang}.} \bibinfo{year}{2001}\natexlab{}.
\newblock \bibinfo{booktitle}{\emph{Accessing and browsing information and communication}}.
\newblock \bibinfo{publisher}{MIT Press}.
\newblock


\bibitem[Romero and Oh(2024)]%
        {RomeroWoven2024}
\bibfield{author}{\bibinfo{person}{Lisette~A Romero} {and} \bibinfo{person}{Hyunjoo Oh}.} \bibinfo{year}{2024}\natexlab{}.
\newblock \showarticletitle{Woven Speaker: Exploring the Potential of Craft as a Method for Exploring Technology through Weaving}. In \bibinfo{booktitle}{\emph{Companion Publication of the 2024 ACM Designing Interactive Systems Conference}} (IT University of Copenhagen, Denmark) \emph{(\bibinfo{series}{DIS '24 Companion})}. \bibinfo{publisher}{Association for Computing Machinery}, \bibinfo{address}{New York, NY, USA}, \bibinfo{pages}{314–317}.
\newblock
\showISBNx{9798400706325}
\href{https://doi.org/10.1145/3656156.3665426}{doi:\nolinkurl{10.1145/3656156.3665426}}


\bibitem[Rosselli Del~Turco and Dalsgaard(2024)]%
        {Rosselli2024Ideas}
\bibfield{author}{\bibinfo{person}{Emilia Rosselli Del~Turco} {and} \bibinfo{person}{Peter Dalsgaard}.} \bibinfo{year}{2024}\natexlab{}.
\newblock \showarticletitle{“My ideas come little by little”: how graphic professionals manage ideas}. In \bibinfo{booktitle}{\emph{Proceedings of the 16th Conference on Creativity \& Cognition}} (Chicago, IL, USA) \emph{(\bibinfo{series}{C\&C '24})}. \bibinfo{publisher}{Association for Computing Machinery}, \bibinfo{address}{New York, NY, USA}, \bibinfo{pages}{452–463}.
\newblock
\showISBNx{9798400704857}
\href{https://doi.org/10.1145/3635636.3656203}{doi:\nolinkurl{10.1145/3635636.3656203}}


\bibitem[Rudnicki(2021)]%
        {rudnicki2021ideas}
\bibfield{author}{\bibinfo{person}{Seweryn Rudnicki}.} \bibinfo{year}{2021}\natexlab{}.
\newblock \showarticletitle{What are ideas made of? On the socio-materiality of creative processes}.
\newblock \bibinfo{journal}{\emph{Creativity studies}} \bibinfo{volume}{14}, \bibinfo{number}{1} (\bibinfo{year}{2021}), \bibinfo{pages}{187--196}.
\newblock


\bibitem[Semaan(2019)]%
        {Semaan2019}
\bibfield{author}{\bibinfo{person}{Bryan Semaan}.} \bibinfo{year}{2019}\natexlab{}.
\newblock \showarticletitle{'Routine Infrastructuring' as 'Building Everyday Resilience with Technology': When Disruption Becomes Ordinary}.
\newblock \bibinfo{journal}{\emph{Proc. ACM Hum.-Comput. Interact.}} \bibinfo{volume}{3}, \bibinfo{number}{CSCW}, Article \bibinfo{articleno}{73} (\bibinfo{date}{Nov.} \bibinfo{year}{2019}), \bibinfo{numpages}{24}~pages.
\newblock
\href{https://doi.org/10.1145/3359175}{doi:\nolinkurl{10.1145/3359175}}


\bibitem[Shelby et~al\mbox{.}(2024)]%
        {shelby2024creative}
\bibfield{author}{\bibinfo{person}{Renee Shelby}, \bibinfo{person}{Ramya Srinivasan}, \bibinfo{person}{Katharina Burgdorf}, \bibinfo{person}{Jennifer~C Lena}, {and} \bibinfo{person}{Negar Rostamzadeh}.} \bibinfo{year}{2024}\natexlab{}.
\newblock \showarticletitle{Creative ML Assemblages: The Interactive Politics of People, Processes, and Products}.
\newblock \bibinfo{journal}{\emph{Proceedings of the ACM on Human-Computer Interaction}} \bibinfo{volume}{8}, \bibinfo{number}{CSCW1} (\bibinfo{year}{2024}), \bibinfo{pages}{1--30}.
\newblock


\bibitem[Simpson et~al\mbox{.}(2023)]%
        {simpson2023captions}
\bibfield{author}{\bibinfo{person}{Ellen Simpson}, \bibinfo{person}{Samantha Dalal}, {and} \bibinfo{person}{Bryan Semaan}.} \bibinfo{year}{2023}\natexlab{}.
\newblock \showarticletitle{"Hey, Can You Add Captions?": The Critical Infrastructuring Practices of Neurodiverse People on TikTok}.
\newblock \bibinfo{journal}{\emph{Proceedings of the ACM on Human-Computer Interaction}} \bibinfo{volume}{7}, \bibinfo{number}{CSCW1} (\bibinfo{year}{2023}), \bibinfo{pages}{1--27}.
\newblock


\bibitem[Simpson et~al\mbox{.}(2022)]%
        {simpson2022tame}
\bibfield{author}{\bibinfo{person}{Ellen Simpson}, \bibinfo{person}{Andrew Hamann}, {and} \bibinfo{person}{Bryan Semaan}.} \bibinfo{year}{2022}\natexlab{}.
\newblock \showarticletitle{How to tame" your" algorithm: LGBTQ+ users' domestication of TikTok}.
\newblock \bibinfo{journal}{\emph{Proceedings of the ACM on Human-computer Interaction}} \bibinfo{volume}{6}, \bibinfo{number}{GROUP} (\bibinfo{year}{2022}), \bibinfo{pages}{1--27}.
\newblock


\bibitem[Simpson and Semaan(2021)]%
        {foryouforyou}
\bibfield{author}{\bibinfo{person}{Ellen Simpson} {and} \bibinfo{person}{Bryan Semaan}.} \bibinfo{year}{2021}\natexlab{}.
\newblock \showarticletitle{For You, or For"You"? Everyday LGBTQ+ Encounters with TikTok}.
\newblock \bibinfo{journal}{\emph{Proc. ACM Hum.-Comput. Interact.}} \bibinfo{volume}{4}, \bibinfo{number}{CSCW3}, Article \bibinfo{articleno}{252} (\bibinfo{date}{Jan.} \bibinfo{year}{2021}), \bibinfo{numpages}{34}~pages.
\newblock
\href{https://doi.org/10.1145/3432951}{doi:\nolinkurl{10.1145/3432951}}


\bibitem[Simpson and Semaan(2023)]%
        {simpson2023rethinking}
\bibfield{author}{\bibinfo{person}{Ellen Simpson} {and} \bibinfo{person}{Bryan Semaan}.} \bibinfo{year}{2023}\natexlab{}.
\newblock \showarticletitle{Rethinking creative labor: A sociotechnical examination of creativity \& creative work on TikTok}. In \bibinfo{booktitle}{\emph{Proceedings of the 2023 CHI Conference on Human Factors in Computing Systems}}. \bibinfo{pages}{1--16}.
\newblock


\bibitem[Sonenshein(2016)]%
        {sonenshein2016routines}
\bibfield{author}{\bibinfo{person}{Scott Sonenshein}.} \bibinfo{year}{2016}\natexlab{}.
\newblock \showarticletitle{Routines and creativity: From dualism to duality}.
\newblock \bibinfo{journal}{\emph{Organization Science}} \bibinfo{volume}{27}, \bibinfo{number}{3} (\bibinfo{year}{2016}), \bibinfo{pages}{739--758}.
\newblock


\bibitem[Star and Ruhleder(1996)]%
        {star1996steps}
\bibfield{author}{\bibinfo{person}{Susan~Leigh Star} {and} \bibinfo{person}{Karen Ruhleder}.} \bibinfo{year}{1996}\natexlab{}.
\newblock \showarticletitle{Steps toward an ecology of infrastructure: Design and access for large information spaces}.
\newblock \bibinfo{journal}{\emph{Information systems research}} \bibinfo{volume}{7}, \bibinfo{number}{1} (\bibinfo{year}{1996}), \bibinfo{pages}{111--134}.
\newblock


\bibitem[Strauss and Corbin(1990)]%
        {strauss1990basics}
\bibfield{author}{\bibinfo{person}{Anselm Strauss} {and} \bibinfo{person}{Juliet Corbin}.} \bibinfo{year}{1990}\natexlab{}.
\newblock \bibinfo{booktitle}{\emph{Basics of qualitative research}}.
\newblock \bibinfo{publisher}{Sage Publications}.
\newblock


\bibitem[Thrash et~al\mbox{.}(2014)]%
        {thrash2014psychology}
\bibfield{author}{\bibinfo{person}{Todd~M Thrash}, \bibinfo{person}{Emil~G Moldovan}, \bibinfo{person}{Victoria~C Oleynick}, {and} \bibinfo{person}{Laura~A Maruskin}.} \bibinfo{year}{2014}\natexlab{}.
\newblock \showarticletitle{The psychology of inspiration}.
\newblock \bibinfo{journal}{\emph{Social and personality psychology compass}} \bibinfo{volume}{8}, \bibinfo{number}{9} (\bibinfo{year}{2014}), \bibinfo{pages}{495--510}.
\newblock


\bibitem[Ungless et~al\mbox{.}(2024)]%
        {ungless2024experiences}
\bibfield{author}{\bibinfo{person}{Eddie~L Ungless}, \bibinfo{person}{Nina Markl}, {and} \bibinfo{person}{Bj{\"o}rn Ross}.} \bibinfo{year}{2024}\natexlab{}.
\newblock \showarticletitle{Experiences of censorship on TikTok across marginalised identities}.
\newblock \bibinfo{journal}{\emph{arXiv preprint arXiv:2407.14164}} (\bibinfo{year}{2024}).
\newblock


\bibitem[Vyas(2019)]%
        {Vyasaltrusism2019}
\bibfield{author}{\bibinfo{person}{Dhaval Vyas}.} \bibinfo{year}{2019}\natexlab{}.
\newblock \showarticletitle{Altruism and Wellbeing as Care Work in a Craft-based Maker Culture}.
\newblock \bibinfo{journal}{\emph{Proc. ACM Hum.-Comput. Interact.}} \bibinfo{volume}{3}, \bibinfo{number}{GROUP}, Article \bibinfo{articleno}{239} (\bibinfo{date}{dec} \bibinfo{year}{2019}), \bibinfo{numpages}{12}~pages.
\newblock
\href{https://doi.org/10.1145/3361120}{doi:\nolinkurl{10.1145/3361120}}


\bibitem[Wallace(1987)]%
        {wallace1987using}
\bibfield{author}{\bibinfo{person}{David~L Wallace}.} \bibinfo{year}{1987}\natexlab{}.
\newblock \showarticletitle{Using peer tutors to overcome writer's block}.
\newblock \bibinfo{journal}{\emph{Research and Teaching in Developmental Education}} \bibinfo{volume}{3}, \bibinfo{number}{2} (\bibinfo{year}{1987}), \bibinfo{pages}{32--41}.
\newblock


\bibitem[Wan and Lu(2023)]%
        {WanIdeation2023}
\bibfield{author}{\bibinfo{person}{Qian Wan} {and} \bibinfo{person}{Zhicong Lu}.} \bibinfo{year}{2023}\natexlab{}.
\newblock \showarticletitle{GANCollage: A GAN-Driven Digital Mood Board to Facilitate Ideation in Creativity Support}. In \bibinfo{booktitle}{\emph{Proceedings of the 2023 ACM Designing Interactive Systems Conference}} (Pittsburgh, PA, USA) \emph{(\bibinfo{series}{DIS '23})}. \bibinfo{publisher}{Association for Computing Machinery}, \bibinfo{address}{New York, NY, USA}, \bibinfo{pages}{136–146}.
\newblock
\showISBNx{9781450398930}
\href{https://doi.org/10.1145/3563657.3596072}{doi:\nolinkurl{10.1145/3563657.3596072}}


\bibitem[Weber(1919)]%
        {weber1919science}
\bibfield{author}{\bibinfo{person}{M Weber}.} \bibinfo{year}{1919}\natexlab{}.
\newblock \showarticletitle{Science as a Vocation [in:] HH Gerth, C. Wright Mills (1946)}.
\newblock \bibinfo{journal}{\emph{From Max Weber: Essays in Sociology}} (\bibinfo{year}{1919}).
\newblock


\bibitem[Wengraf(2001)]%
        {wengraf2001qualitative}
\bibfield{author}{\bibinfo{person}{Tom Wengraf}.} \bibinfo{year}{2001}\natexlab{}.
\newblock \bibinfo{booktitle}{\emph{Qualitative research interviewing: Biographic narrative and semi-structured methods}}.
\newblock \bibinfo{publisher}{Sage Publications Sage CA: Los Angeles, CA}.
\newblock


\bibitem[Wheeldon and Faubert(2009)]%
        {wheeldon2009framing}
\bibfield{author}{\bibinfo{person}{Johannes Wheeldon} {and} \bibinfo{person}{Jacqueline Faubert}.} \bibinfo{year}{2009}\natexlab{}.
\newblock \showarticletitle{Framing experience: Concept maps, mind maps, and data collection in qualitative research}.
\newblock \bibinfo{journal}{\emph{International journal of qualitative methods}} \bibinfo{volume}{8}, \bibinfo{number}{3} (\bibinfo{year}{2009}), \bibinfo{pages}{68--83}.
\newblock


\end{thebibliography}

\clearpage
\appendix
\section{Appendix A: Participant Demographics Table}
\begin{table}[h]
\centering
\begin{tabular}{|c|c|l|c|l|c|l|}
\hline
\textbf{\#} & \textbf{Age} & \textbf{Gender}                                              	& \textbf{Pronouns} & \textbf{Race/Ethnicity}                                                  	& \textbf{Locale} & \textbf{Art They Do}                                                                           	\\ \hline
1       	& 31       	& Non-Binary                                                   	& she/they      	& Black                                                                    	& Urban       	& Illustrator                                                                            	\\ \hline
2       	& 21       	& \begin{tabular}[c]{@{}l@{}}Nonbinary /\\ Transmasc\end{tabular}  & they/them     	& White                                                                    	& Urban       	& Animator / Artist                                                                      	\\ \hline
3       	& 28       	& \begin{tabular}[c]{@{}l@{}}nonbinary, \\ trans\end{tabular}  	& they/them     	& White, Ashkenazi                                                         	& Urban       	& \begin{tabular}[c]{@{}l@{}}Freelance Artist / Digital \\ Artist / Illustrator\end{tabular} \\ \hline
4       	& 29       	& \begin{tabular}[c]{@{}l@{}}Non-binary / \\ Demigirl\end{tabular} & she/they      	& \begin{tabular}[c]{@{}l@{}}Mixed (White \& \\ African American)\end{tabular} & Suburban    	& Illustrator                                                                            	\\ \hline
5       	& 21       	& woman                                                        	& she/Her       	& Hispanic/Latinx                                                          	& Suburban    	& Artist                                                                                 	\\ \hline
6       	& 28       	& cis woman                                                    	& she/Her       	& White British                                                            	& Rural       	& Comic Artist / Illustrator                                                             	\\ \hline
7       	& 65       	& Male                                                         	& he/him        	& White/Caucasian                                                          	& Urban       	& Sculptor (In Wood)                                                                      	\\ \hline
8       	& 32       	& Male                                                         	& he/him        	& White                                                                    	& Rural       	& Wood Intarsia                                                                          	\\ \hline
9       	& 33       	& non-binary                                                   	& they/them     	& \begin{tabular}[c]{@{}l@{}}White (Eastern \\ European)  \end{tabular}                                               	& Rural       	& Comic Artist / Illustrator                                                             	\\ \hline
10      	& 28       	& Cis Woman                                                    	& she/her       	& Latina                                                                   	& Urban       	& Hobby Artist                                                                           	\\ \hline
11      	& 74       	& Male                                                         	& he/him        	& White                                                                    	& Suburban    	& \begin{tabular}[c]{@{}l@{}}Abstract Color Painting \\ (mixed media artist)\end{tabular}	\\ \hline
12      	& 27       	& Butch                                                        	& she/her       	& White                                                                    	& Suburban    	& Doodler                                                                                	\\ \hline
13      	& 33       	& Woman                                                        	& she/Her       	& Caucasian                                                                	& Rural       	& Photographer / Crafter                                                                 	\\ \hline
14      	& 25       	& Trans woman                                                  	& she/Her       	& Caucasian                                                                	& Suburban    	& Multi-Media Artist                                                                     	\\ \hline
15      	& 30       	& Woman                                                        	& she/her       	& Latina                                                                   	& Suburban    	& Digital Artist / Illustrator                                                           	\\ \hline
16      	& 24       	& Male                                                         	& he/him        	& White                                                                    	& Rural       	& Epoxy Resin / Floral Presser                                                           	\\ \hline
17      	& 29       	& Male                                                         	& he/him        	& White/Mexican                                                            	& Urban       	& Digital Artist                                                                         	\\ \hline
18      	& 54       	& Male                                                         	& he/him        	& Caucasian                                                                	& Urban       	& Storyteller, Comic Books                                                               	\\ \hline

19      	& 68       	& Woman                                                        	& she/her       	& White                                                                  	& Urban    	& Bookmaker                                                          	\\ \hline
20      	& 34       	& Male                                                         	& he/him        	& White                                                                    	& Suburban       	& \begin{tabular}[c]{@{}l@{}}Found \& Recycled Materials\\ Instrument Builder (Luthier)\end{tabular}                                                       	\\ \hline
21      	& 18       	& Refused                                                         	& she/her        	& Asian American                                                        	& Urban       	& \begin{tabular}[c]{@{}l@{}}Mixed Media / Fiber Artist\\ / Photography  \end{tabular}                                                                        	\\ \hline
22      	& 37       	&Female                                                        	& she/her        	& Caucasian                                                                	& Rural       	& Multi-Media Artist \\ \hline
\end{tabular}
\caption{Participant Demographics, as they described themselves}
\Description[Participant Demographics, as they described themselves.]{Participant Demographics, as they described themselves.}
\label{tab:participants}
\end{table}

\end{document}